\documentstyle[rotating,epsfig,subfigure]{elsart}

\def\boldsymbol{\bf}

\input epsf 

\def\dx{\ifmmode {\ \mathrm{d} x} \else $\mathrm{d} x$ \fi}
\def\ln{\ifmmode {\ \mathrm{ln}} \else $\mathrm{ln}$ \fi}
\def\min{\ifmmode {\ \mathrm{min}} \else $\mathrm{min}$ \fi}
\def\d{\ifmmode {\ \mathrm{d}} \else $\mathrm{d}$ \fi}
\def\Sr{\ifmmode {\ ^{90}\mathrm{Sr}} \else $^{90}\mathrm{Sr}$ \fi}
\def\Y{\ifmmode {\ ^{90}\mathrm{Y}} \else $^{90}\mathrm{Y}$ \fi}
\def\Zr{\ifmmode {\ ^{90}\mathrm{Zr}} \else $^{90}\mathrm{Zr}$ \fi}
\def\Cm{\ifmmode {\ ^{244}\mathrm{Cm}} \else $^{244}\mathrm{Cm}$ \fi}
\def\Am{\ifmmode {\ ^{241}\mathrm{Am}} \else $^{241}\mathrm{Am}$ \fi}
\def\muCi{\ifmmode {\ \mathrm{\mu Ci}} \else $\mathrm{\mu Ci}$ \fi}
\def\mCi{\ifmmode {\ \mathrm{mCi}} \else $\mathrm{mCi}$ \fi}
\def\rad{\ifmmode {\ \mathrm{rad}} \else $\mathrm{rad}$ \fi}
\def\krad{\ifmmode {\ \mathrm{krad}} \else $\mathrm{krad}$ \fi}

\def\K{\ifmmode {\ \mathrm{K}} \else $\mathrm{K}$ \fi} \def\h{\ifmmode
{\ \mathrm{h}} \else $\mathrm{h}$ \fi} \def\s{\ifmmode {\ \mathrm{s}}
\else $\mathrm{s}$ \fi} \def\ms{\ifmmode {\ \mathrm{ms}} \else
$\mathrm{ms}$ \fi} \def\mus{\ifmmode {\ \mathrm{\mu s}} \else
$\mathrm{\mu s}$ \fi} \def\ns{\ifmmode {\ \mathrm{ns}} \else
$\mathrm{ns}$ \fi} \def\mm{\ifmmode {\ \mathrm{mm}} \else
$\mathrm{mm}$ \fi} \def\cm{\ifmmode {\ \mathrm{cm}} \else
$\mathrm{cm}$ \fi} \def\m{\ifmmode {\ \mathrm{m}} \else $\mathrm{m}$
\fi} \def\mum{\ifmmode {\ \mathrm{\mu m}} \else $\mathrm{\mu m}$ \fi}
\def\nm{\ifmmode {\ \mathrm{nm}} \else $\mathrm{nm}$ \fi}
\def\kHz{\ifmmode {\ \mathrm{kHz}} \else $\mathrm{kHz}$ \fi}
\def\Hz{\ifmmode {\ \mathrm{Hz}} \else $\mathrm{Hz}$ \fi}
\def\V{\ifmmode {\ \mathrm{V}} \else $\mathrm{V}$ \fi}
\def\kV{\ifmmode {\ \mathrm{kV}} \else $\mathrm{kV}$ \fi}
\def\Ohm{\ifmmode {\ \Omega} \else $\Omega$ \fi} \def\Tohm{\ifmmode {\
\mathrm{T}\Omega} \else $\mathrm{T}\Omega$ \fi} \def\pF{\ifmmode {\
\mathrm{pF}} \else $\mathrm{pF}$ \fi} \def\ENC{\ifmmode {\
\mathrm{ENC}} \else $\mathrm{ENC}$ \fi} \def\e{\ifmmode {\ \mathrm{e}}
\else $\mathrm{e}$ \fi} \def\pC{\ifmmode {\ \mathrm{pC}} \else
$\mathrm{pC}$ \fi} \def\eh{\ifmmode {\ \mathrm{eh}} \else
$\mathrm{eh}$ \fi} \def\n{\ifmmode {\ \mathrm{n}} \else $\mathrm{n}$
\fi} \def\ke{\ifmmode {\ \mathrm{ke}} \else $\mathrm{ke}$ \fi}
\def\GeV{\ifmmode { \mathrm{GeV}} \else $\mathrm{GeV}$ \fi}
\def\GeVc{\ifmmode { \mathrm{GeV/c}} \else $\mathrm{GeV/c}$ \fi}
\def\GeVcc{\ifmmode { \mathrm{GeV/c^2}} \else $\mathrm{GeV/c^2}$ \fi}
\def\MeV{\ifmmode {\ \mathrm{MeV}} \else $\mathrm{MeV}$ \fi}
\def\keV{\ifmmode {\ \mathrm{keV}} \else $\mathrm{keV}$ \fi}
\def\eV{\ifmmode {\ \mathrm{eV}} \else $\mathrm{eV}$ \fi}
\def\MeVc{\ifmmode {\ \mathrm{MeV/c}} \else $\mathrm{MeV/c}$ \fi}
\def\MeVcc{\ifmmode {\ \mathrm{MeV/c^2}} \else $\mathrm{MeV/c^2}$ \fi}
\def\mbar{\ifmmode {\ \mathrm{mbar}} \else $\mathrm{mbar}$ \fi}
\def\ps{\ifmmode {\ \mathrm{ps}} \else ps \fi} \def\pb{ {\ifmmode
\;{{\mbox{\mathrm pb}^{-1}}} \else { pb$^{-1}$ } \fi }}
\def\dEdx{\ifmmode {{\mathrm dE}{\mathrm dx}} \else {${\mathrm
dE}{\mathrm dx}$} \fi } \def\Torr{\ifmmode {\ \mathrm{Torr}} \else
$\mathrm{Torr}$ \fi} \def\bunch{\ifmmode {\ \mathrm{bunch}} \else
$\mathrm{bunch}$ \fi} \def\b{\ifmmode {\ \mathrm{b}} \else
$\mathrm{b}$ \fi} \def\C{\ifmmode {\ ^{\circ} \mathrm{C}} \else
$^{\circ} \mathrm{C}$ \fi}

\begin{document}
\epsfverbosetrue

\begin{frontmatter}
\title{The Charge Collection Properties of CVD Diamond }
\author[CERN]{Ties Behnke}, \author[DESY,uni]{Petra H\"untemeyer},
\author[DESY,uni]{Alexander Oh\thanksref{corres}},
\author[DESY,uni]{Johannes Steuerer}, \author[DESY,uni]{Albrecht
Wagner} and \author[DESY]{Wolfram Zeuner} \address[CERN]{CERN PPE,
1211 Geneva 23, Switzerland} \address[uni]{II. Institut f\"ur
Experimentalphysik, Universit\"at Hamburg, Luruper Chaussee 149, DESY
Bldg. 62, Germany} \address[DESY]{DESY, Postfach, 22603 Hamburg,
Germany} \thanks[corres]{Corresponding author, e-mail:
Alexander.Oh@desy.de.}


\begin{abstract}
The charge collection properties of CVD diamond have been investigated
with ionising radiation. In this study two CVD diamond samples,
prepared with electrical contacts have been used as solid state
ionisation chambers.  The diamonds have been studied with beta
particles and 10 keV photons, providing a homogeneous ionisation
density and with protons and alpha particles which are absorbed in a
thin surface layer.  For the latter case a strong decrease of the
signal as function of time is observed, which is attributed to
polarisation effects inside the diamond.  Spatially resolved
measurements with protons show a large variation of the charge
collection efficiency, whereas for photons and minimum ionising
particles the response is much more uniform and in the order of 18\%.
These results indicate that the applicability of CVD diamond as a
position sensitive particle detector depends on the ionisation type
and appears to be promising for homogeneous ionisation densities as
provided by relativistic charged particles.

\end{abstract}
\end{frontmatter}

\section{Introduction}
\label{s-introduction}
In recent years, diamond as a possible material for particle detectors
has been the subject of considerable
interest~\cite{ref-cd,ref-rd42_94,ref-euro96_pochet}.  Significant
progress in the techniques to produce synthetical diamond films of
very high quality has been achieved by means of the Chemical Vapour
Deposition method (CVD).  A number of commercial manufactures of CVD
diamond films~\cite{ref-norton,ref-debeers} and research
institutes~\cite{ref-cd_record} have made systematic studies of the
properties of this material feasible. The main advantage of the
material compared to other semiconductor detector materials is its
radiation hardness, which has recently been demonstrated to neutron
fluences of up to $10^{14}\cm^{-2}$\cite{ref-hassard}. The radiation
hardness of the material is of strong interest for the detector
development at projected experiments where high radiation levels are
expected due to the increasing luminosity and energy, as e.g. the
experiments at the Large Hadron Collider at CERN.  The main problem of
using CVD diamond as a detector material are the charge collection
properties, since an application as a detector for ionising particles
requires that the material response is homogeneous throughout the
volume and that a sufficient fraction of the produced charge is
collected. General studies of the CVD diamond growth conditions for
detector applications with a large sample are presented
in~\cite{ref-cd_record}.

In this paper we present different studies of the charge collection
properties of CVD diamond films. We investigate both the bulk
properties of the material and the behaviour of charge produced close
to the surface.

\subsection{Principle}
\begin{figure}[tmb]
\begin{center}
\mbox{\epsfig{file=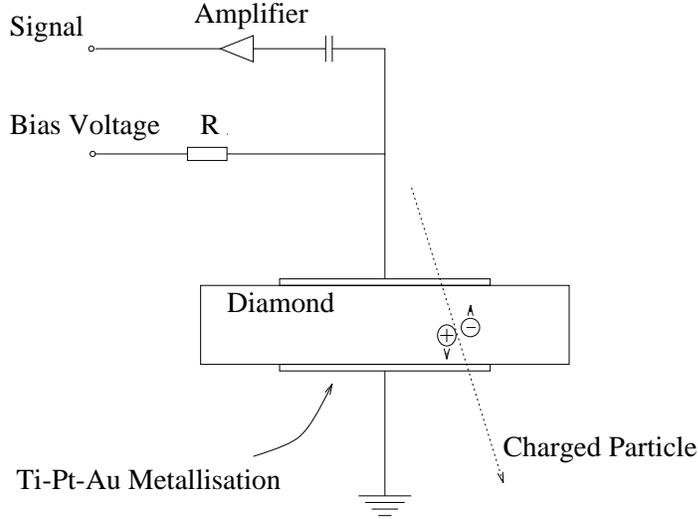,bbllx=0.cm,bblly=0.cm,bburx=6.5in
,bbury=5.in,height=7.cm}}
\caption[]{\label{fig-detector}Principle of a diamond detector.}
\end{center}
\end{figure}

One of the aspects which make CVD diamond films an attractive material
for the detector development is its very high specific resistivity of
$>10^{13} \Ohm\cm$. This allows a very simple construction of a
detector as a solid state ionisation chamber (see
fig.~\ref{fig-detector}). By contacting the material on opposite sides
and applying a sufficient potential across the contacts, charges
produced in the bulk of the material by an incident charged particle
start to drift towards the electrodes
and induce mirror charges.  The ratio of the amount of charge
measured, $Q_{\mathrm{m}}$, to the amount of charge produced,
$Q_{\mathrm{0}}$, is the collection efficiency
\[ \epsilon_{\mathrm{Q}}=Q_{\mathrm{m}} / Q_{\mathrm{0}} [\%].\]
The efficiency is below 100\%, if trap- or recombination-sites in the
material hinder the charge carriers to reach the electrodes, as is the
case for diamond. The mean free path length $\delta$, also called
Schubweg~\cite{ref-hofstadter}, for each type of charge carriers
($c=e,h$ for electrons and holes) can be expressed as the product of
the mobility $\mu_{\mathrm{c}}$, the electric field strength $E$ and
the life-time $\tau_{\mathrm{c}}$, as
\[ \delta_{\mathrm{c}} = \mu_{\mathrm{c}} \tau_{\mathrm{c}} E .\]
The combined mean free paths of electrons and holes,
\begin{equation}\label{eq-1}
\delta = (\mu_{\mathrm{e}} \tau_{\mathrm{e}} + \mu_{\mathrm{h}}
\tau_{\mathrm{h}}) \cdot E ,
\end{equation}
is the mean distance a hole and an electron can separate. This
distance is referred to as the collection distance~\cite{ref-cd}.  The
collection distance and the efficiency $\epsilon_{\mathrm{Q}}$ are
related according to Ramo's theorem~\cite{ref-ramo} by
\begin{equation}\label{eq-cd}
\epsilon_{\mathrm{Q}} \approx {\delta \over d},
\end{equation}
with $d$ being the detector thickness.

The above discussion holds for uniform ionisation within the detector
volume and $\delta \ll d$. If the charge is created in the vicinity of
one electrode ($x \ll d$, with $x$ the distance of the charge to the
electrode) the van Hecht-equation~\cite{ref-hecht} relates
$\epsilon_{\mathrm{Q}}$ and $\delta$ as
\begin{equation}\label{eq-hecht}
\epsilon_{\mathrm{Q}} \approx {\delta_{\mathrm{c}} \over d}
\end{equation}
where $\delta_{\mathrm{c}} \ll d$ denotes the Schubweg for electrons
and holes respectively, depen-\\ding on the direction of the
electrical field.

Throughout this paper, we will use the quantity
$\epsilon_{\mathrm{Q}}$ because the definition of
$\epsilon_{\mathrm{Q}}$ involves fewer assumptions than the charge
collection distance. However, the values are given in collection
distance as well where appropriate.

Systematic studies of the dependence of $\epsilon_{\mathrm{Q}}$ on
growth parameters have shown that high values of
$\epsilon_{\mathrm{Q}}$ are reached for films produced with slow
growth rates~\cite{ref-cd_record}.  Until now charge collection
efficiencies of up to 38\%~\cite{ref-cd_record} have been reported.

It is known that the exposure of diamond to ionising radiation and
UV-light can substantially increase $\epsilon_{\mathrm{Q}}$. This
phenomenon is called priming~\cite{ref-keddy}.  The passivation of
traps by occupation with free charge carriers is the reason for the
observed behaviour. The same mechanism leads to a deterioriation of
the signal as for example seen when irradiating diamond with alpha
particles or protons at energies corresponding to stopping ranges of a
few micron.  The deterioriation is caused by the build-up of space
charge and the resulting compensation of the applied
field~\cite{ref-manfredotti94}.

Another important issue is the homogeneity of $\epsilon_{\mathrm{Q}}$.
It is known that the collection distance (which relates to
$\epsilon_{\mathrm{Q}}$ by eq.~\ref{eq-cd}) varies with the film
thickness, as low values are typically observed on the substrate side
and high values on the growth side. This is due to the fact that
polycrystalline CVD diamond films consist of several columnar
micro-crystallites. Typically the crystallites are very small at the
beginning of the growth process and become larger as the film
thickness increases~\cite{ref-yarbrough}. A linear model has been
proposed to describe the collection distance as a function of the film
thickness~\cite{ref-zhao}.  For an application as a particle detector,
it is important that the average $\epsilon_{\mathrm{Q}}$ is constant
over the active detector area.  While highly energetic minimum
ionising particles yield a rather homogeneous response (see
section~\ref{s-mips}), it has been shown that this is not the case for
particles depositing their charge in a thin surface
layer~\cite{ref-manfredotti94,ref-oh_dipl95}.  Thin films for which
the stopping range of the ionising particle is of the order of the
film thickness seem to exhibit a more homogeneous
response~\cite{ref-jany}.

\subsection{Methods of investigation}
We investigate the charge collection properties of CVD diamonds by
four different methods. The response of CVD diamond to homogeneous
ionisation densities has been measured with beta particles from a \Sr
source. The ionisation density produced by beta particles above
$1.1\MeV$ is similar to the ionisation density produced by minimum
ionising particles (MIPs), which are to be detected in the application
as a vertex detector in a high energy particle physics experiment. By
fitting the pulse height distribution, we obtained information about
the homogeneity of $\epsilon_{\mathrm{Q}}$.  To investigate the
priming effect on the CVD diamond samples, $\epsilon_{\mathrm{Q}}$ has
been measured with beta particles while the samples were under
continuous exposure of ionising radiation.  In a second experiment,
the diamond has been scanned with a beam of 10 keV photons. The
photons ionise via the photo effect and produce when averaged over
many events a similar ionisation density distribution as MIPs.  With
this method the local variations of $\epsilon_{\mathrm{Q}}$ have been
measured with a resolution of about $100\mum$. The results are
compared with the data obtained with MIPs.

In contrast, the other two methods described in this paper use
radiation which produces charge carriers only in a shallow layer below
the surface. Hence, these methods are not measurements of the bulk
properties as are the first two. Since all the produced charge is
deposited in a surface layer of the detector, they are more sensitive
to polarisation effects, as will be discussed in
section~\ref{s-conclusion}.  One method uses alpha particles from an
\Cm source. Various pulse height spectra have been recorded at
different electrical field strengths and polarities in order to
extract information about $\epsilon_{\mathrm{Q}}$. The other method
uses protons with an energy of 2 MeV and allows to do a spatially
resolved measurement.  The recorded pulse spectra are mapped onto the
diamond surface. The spatial information has been used to compare the
data to SEM (scanning-electron-microscopy) pictures and to data
obtained with the 10 keV photon beam described above.

\section{Sample preparation}
\label{s-samples}
For the studies presented here, two CVD diamond samples produced by
different manufacturers (sample A from Norton~\cite{ref-norton_addr}
and sample B from FhG~\cite{ref-fhg_addr}) have been used.  The sample
A was initially grown to a thickness of $500\mum$ after which
$200\mum$ were removed from the substrate side~\cite{ref-norton}.  The
growth side has been polished to reduce the surface roughness, which
has been measured to be smaller than $20\mum$ after
polishing~\cite{ref-norton}. The grain-size on the growth side is in
the order of $100\mum$. No information is available about the grain
size on the substrate side.  The sample size is $10\mm \times 10\mm$.
Sample B has been grown to a thickness of $123\mum$ and is not
polished. The grains have a typical size of $40\mum$ on the growth
side. From SEM we estimate a surface roughness of below $30\mum$. The
sample has a circular shape and a diameter of $14\mm$.  Pole figure
measurements~\cite{ref-polefig} showed no preferred orientation for
sample A and a slight (111)-texture for sample B.

In order to provide electrical contacts, both samples have been
metallised on growth and substrate side with layers of Ti (50~nm), Pt
(30~nm) and Au (60~nm), followed by an annealing step and treatment in
oxygen plasma to reduce surface conductivity~\cite{ref-fhg_metal}.
The quality of the contacts has been checked by measuring the
current-voltage (I-V) characteristic in darkness at room
temperature. All I-V-curves are symmetric, i.e.~independent of
polarity, indicating, that front and back-contact are of the same
type. The specific resistivities and some of the properties of the
samples are listed in table~\ref{tab-samples}.
\begin{table}[h!]
\begin{center}
\begin{tabular}{|l|l|l|l|l|l|l|l|l|}     \hline
Producer & Name & Process & $v_{\mathrm{g}}[\mum/ \h]$ & $d$ [\mum] &
$R$ $[\Ohm \cm]$ \\ \hline \hline Norton &A &DCAJ &no information &300
& $7\times 10^{13}$ \\ \hline FhG &B &MWP & 1.3 &123 & $6.4\times
10^{14}$ \\ \hline
\end{tabular}
\end{center}
\caption[]{\label{tab-samples}Properties of the investigated samples.
  $v_{\mathrm{g}}$ stands for growth rate, $d$ for film thickness and
  $R$ for the specific resistivity. DCAJ stands for {\em direct
  current arc jet process} and MWP for {\em micro wave plasma
  process}.}
\end{table}

\section{Measurements with beta particles}
\label{s-mips}
The average charge collection properties of the diamond samples have
been investigated using beta particles. Since they traverse the
detectors, they probe $\epsilon_{\mathrm{Q}}$ along the particle
trajectory, and result in a measurement averaged over the detector
thickness.  This method is widely used to characterise diamond
samples, since it corresponds most closely to the situation of a real
detector application.

The field dependence of $\epsilon_{\mathrm{Q}}$ has been measured for
the two samples.  From the shape of the distribution of the collected
charge we have obtained information about the variation of
$\epsilon_{\mathrm{Q}}$ laterally.

The dependence of $\epsilon_{\mathrm{Q}}$ on the absorbed radiation
dose of the samples has also been studied.  The priming effect of
ionising radiation has been established, and for each sample the
characteristic priming dose constant $\Phi_{\mathrm{0}}$ has been
derived.

\subsection{Experimental set-up}
\begin{figure}[tmb]
\begin{center}
\mbox{\epsfig{file=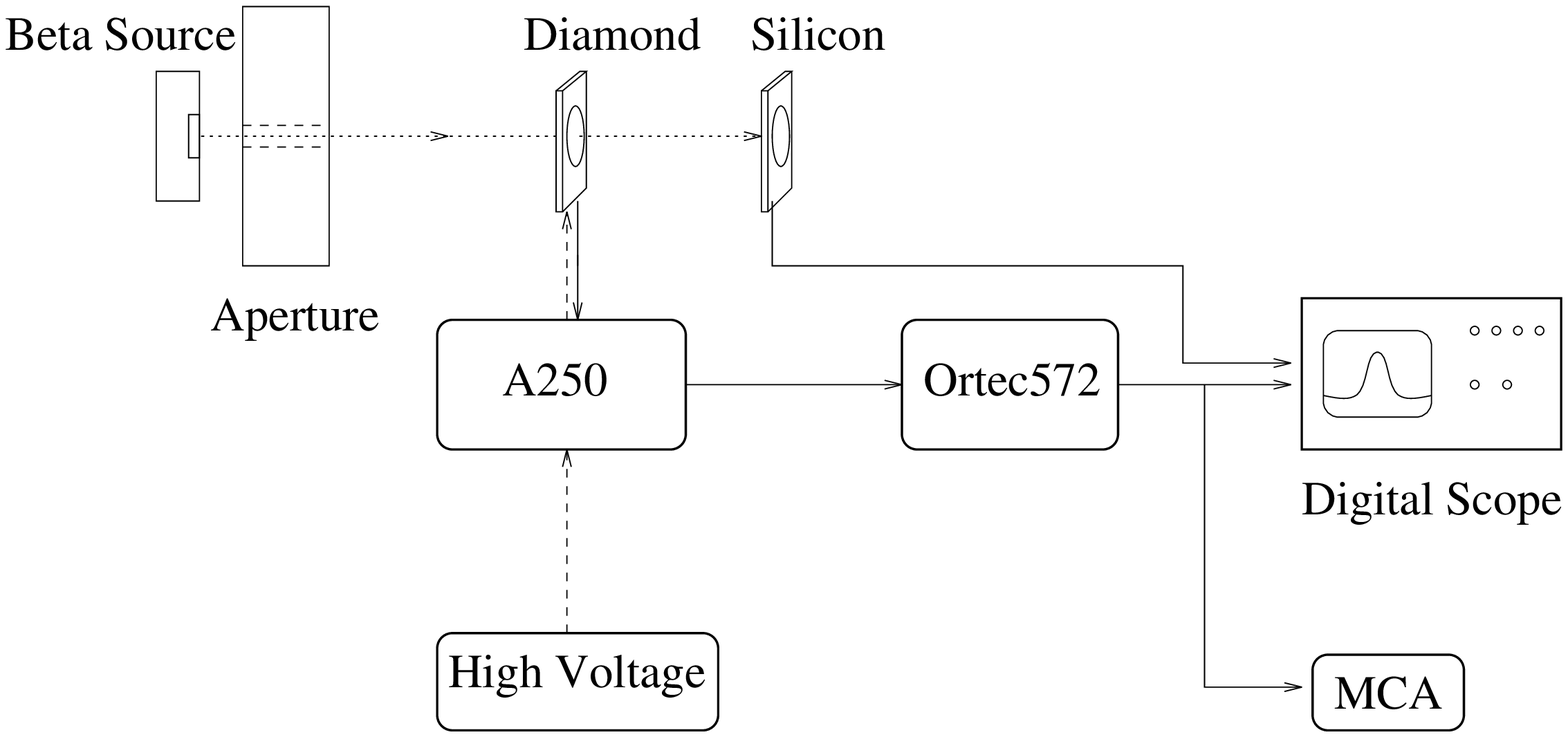,bbllx=0cm,bblly=0.cm
,bburx=8.9in,bbury=4.7in,width=.8\textwidth}}
\caption[]{\label{fig-setup-sr90}Schematic view of the set-up 
for measurements with beta particles.}
\end{center}
\end{figure}

Beta particles have been used emitted from a \Sr source ($114\muCi$)
which have a maximum energy of 2.28 MeV to measure
$\epsilon_{\mathrm{Q}}$ averaged over the bulk of the material.  A
diagram of the experimental set-up is shown in
figure~\ref{fig-setup-sr90}. The beta particles were collimated by a
$1\mm$ aperture in a Plexiglas block. The diamond was placed in the
collimated beam of beta particles followed by a silicon detector which
was used as a trigger device.
The aperture ensured that all beta particles traversed the sensitive
region of the diamond-detector. After having traversed the diamond,
the beta particles had to deposit enough energy ($>100\keV$) in the
silicon-detector to be accepted as an event. The trigger-rate was
limited by the DAQ system to about 15~Hz.  The characteristic
energy-loss of the beta particles in the diamond is approximately that
of minimum ionising particles (MIPs) which on average produces
$\rho_{\mathrm{MIP}}= 36$~electron-hole-pairs~$\mum^{-1}$ in
diamond~\cite{ref-cd}. The collection efficiency
$\epsilon_{\mathrm{Q}}$ is given by
\[\epsilon_{\mathrm{Q}} = {Q_{\mathrm{m}} \over \rho_{\mathrm{MIP}} 
\cdot d },\]
where $d$ is the diamond thickness in $\mum$ and $Q_{\mathrm{m}}$ the
average image charge induced on the electrodes of the diamond
detector.

The charge was measured with a charge-integrating amplifier Amptek
A250 followed by a shaping-amplifier Ortec 572 with a shaping-time
constant of $\tau = 3\mus$. The signal was read out with a digital
scope and averaged over 3000 events to improve the signal-to-noise
ratio.  Additionally, the signal was recorded with a multi channel
analyser (MCA).

In a second experiment the sample was alternately exposed to a strong
\Sr source ($1300\muCi$) without applying a voltage and to the weaker
\Sr source described above to measure $\epsilon_{\mathrm{Q}}$.  With
this set-up the dose dependence of $\epsilon_{\mathrm{Q}}$ was
measured.

\subsection{Experimental results}
\begin{figure}[tmb]
 \begin{center}
  \begin{tabular}{cc}
 \subfigure[] {\epsfig{figure=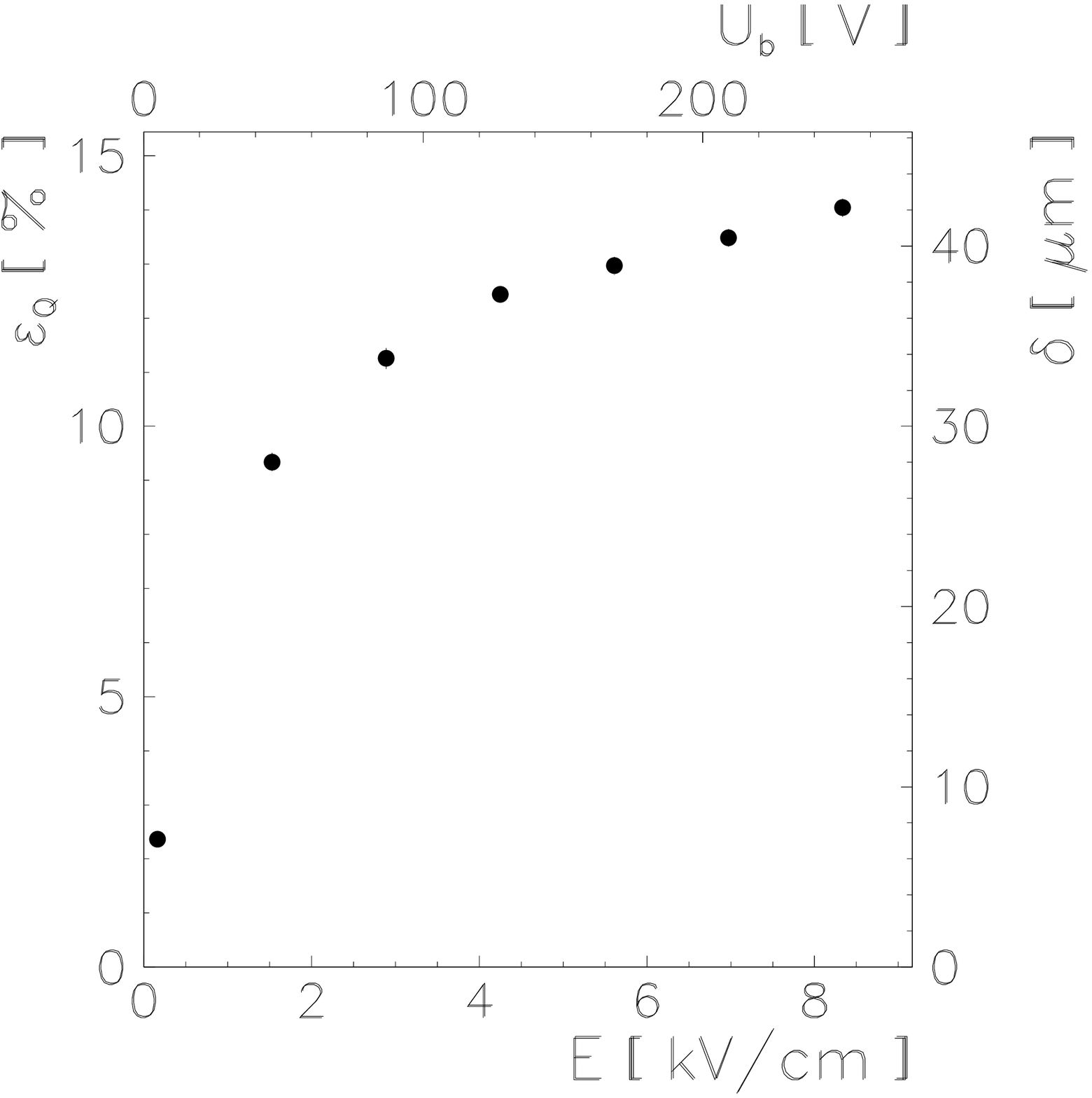,bbllx=0.cm,bblly=0.cm
,bburx=20cm,bbury=20cm,width=0.47\textwidth}} & \subfigure[ ]
{\epsfig{figure=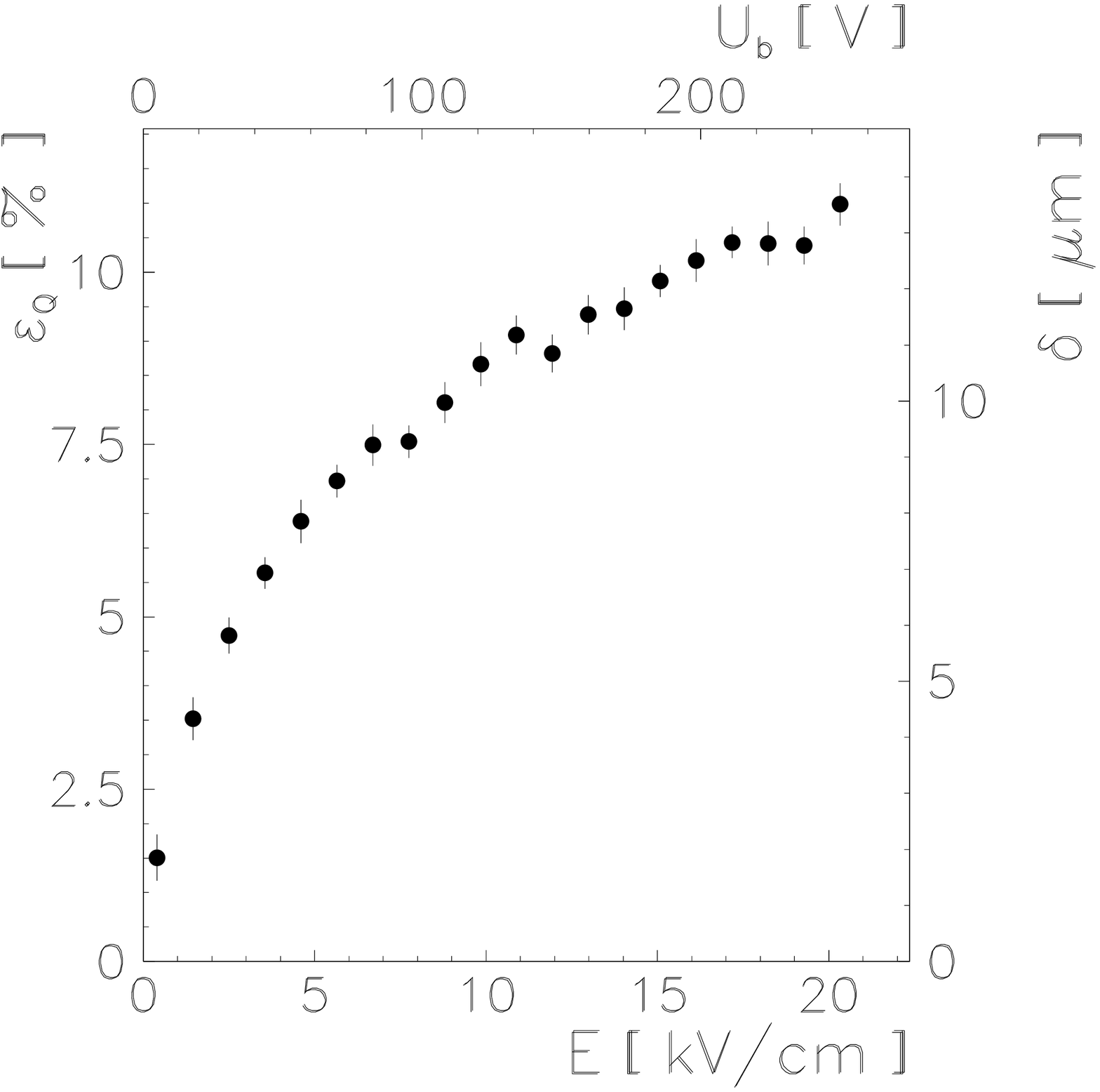,bbllx=0.cm,bblly=0.cm
,bburx=20.cm,bbury=20.cm,width=0.47\textwidth}}
  \end{tabular}
 \end{center}
\caption[]{\label{fig-cd_E}The development of $\epsilon_{\mathrm{Q}}$
  as a function of the electric field for unprimed sample A (a) and
  sample B (b). The equivalent collection distance is given on the
  right scale.}
\end{figure}
The charge collection efficiency $\epsilon_{\mathrm{Q}}$ as a function
of the field strength $E$ (fig.~\ref{fig-cd_E}) exhibits a sharp rise
with $E$ up to about 1~kV/cm for both samples. At higher field
strengths, the curve flattens due to enhanced phonon creation and the
resulting decrease of the mobility~\cite{ref-cd}. Above an applied
voltage of $U_{\mathrm{b}} = 250\V$, the signal became noisy for both
samples, because the leakage current increased. As the corresponding
field strengths differ by a factor $2.4$ this is taken as indication
for surface currents.

\begin{figure}[tmb]
 \begin{center}
  \begin{tabular}{cc}
 \subfigure[ ] {\epsfig{figure=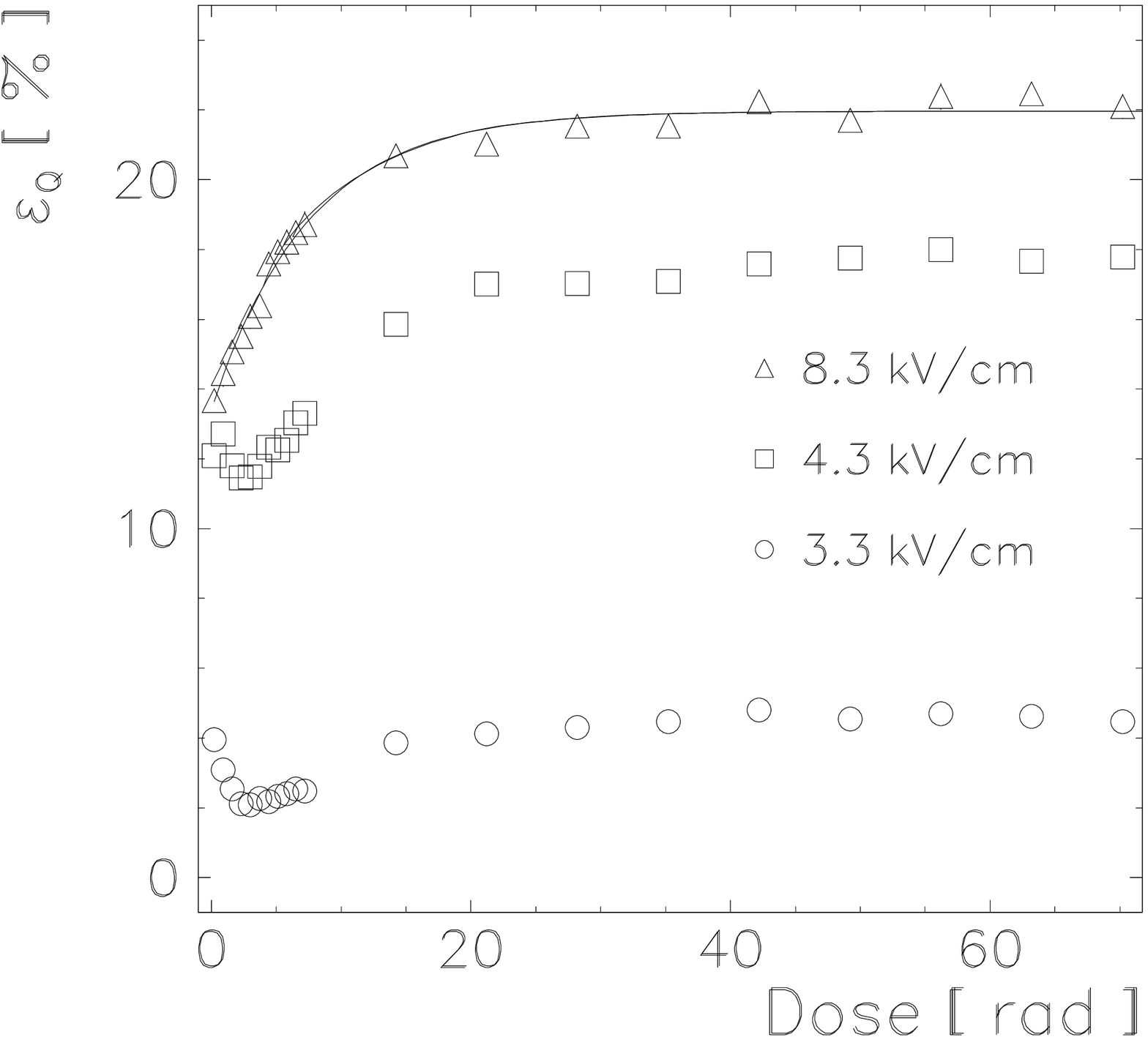,bbllx=0.cm,bblly=0.cm
,bburx=20cm,bbury=20cm,width=0.47\textwidth}} & \subfigure[ ]
{\epsfig{figure=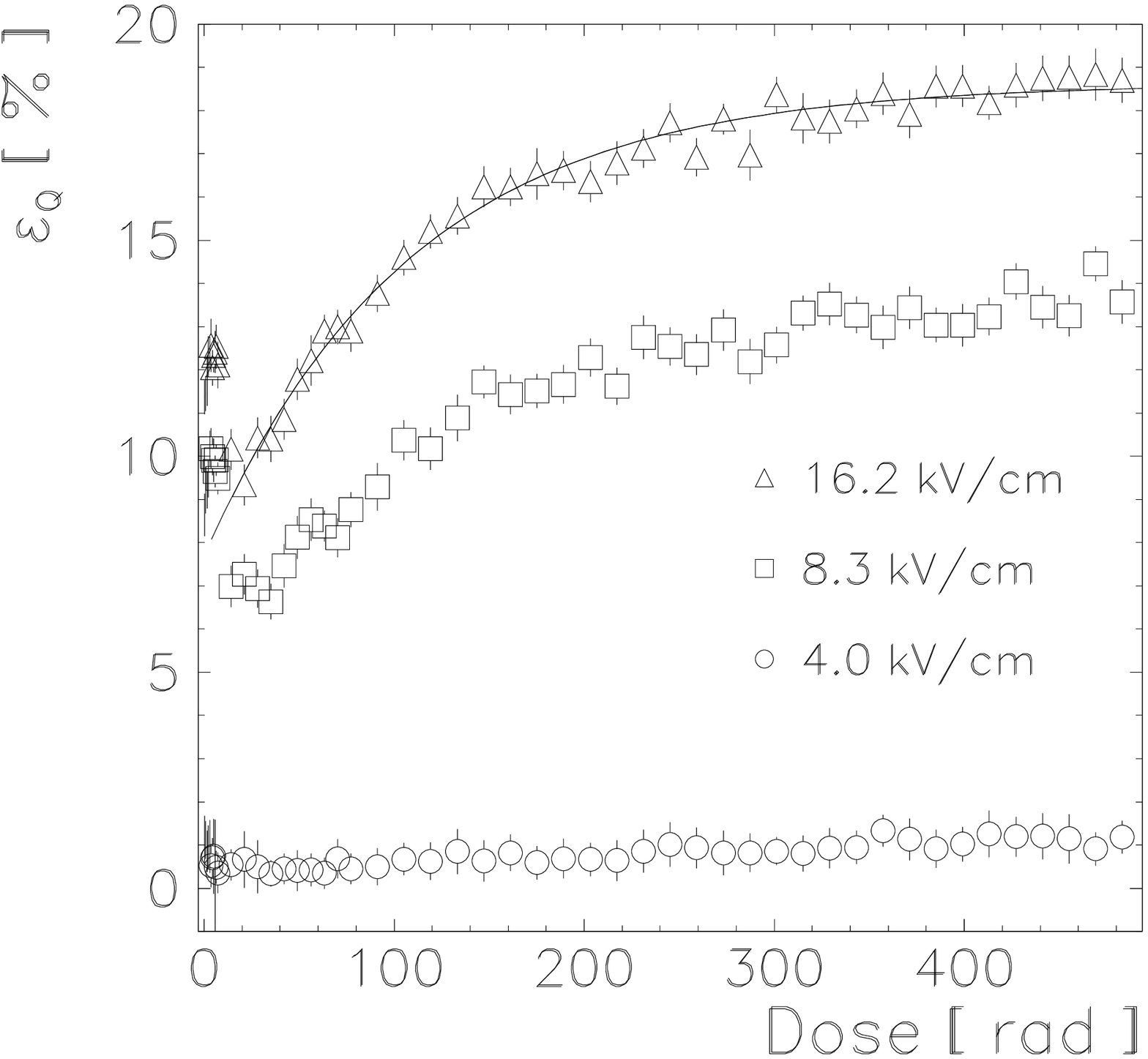,bbllx=0.cm,bblly=0.cm
,bburx=20.cm,bbury=20.cm,width=0.47\textwidth}}
  \end{tabular}
 \end{center}
\caption[]{\label{fig-dose_dep}The dose dependent development of the
charge collection efficiency for sample A (a) and sample B (b) at
different field strengths. Drawn is also the fit
function~\ref{eq-prime-fit}.}
\end{figure}

When the samples were irradiated, a dose-dependent increase of
$\epsilon_{\mathrm{Q}}$ was observed.  In fig. ~\ref{fig-dose_dep},
the dose-dependent development of $\epsilon_{\mathrm{Q}}$ for three
different field strengths is shown.  The initial drop of
$\epsilon_{\mathrm{Q}}$ on sample B is suspected to be due to
polarisation effects. The same behaviour has been observed for sample
A at 4.3 kV/cm and 3.3 kV/cm.  We have fitted a function of the form
\begin{equation} \label{eq-prime-fit}
\delta (\Phi) = a \cdot (1-\e^{- {\Phi \over \Phi_{\mathrm{0}}}}) + b
\end{equation}
to the monotonically rising part of the data and extracted the {\em
priming-dose constant} $\Phi_{\mathrm{0}}$.

The unprimed and primed efficiencies, the ratio of unprimed over
primed efficiency, and $\Phi_{\mathrm{0}}$ for the two samples are
summarised in tab.~\ref{tab-dose_dep}.  The ratio of
$\epsilon_{\mathrm{Q}}$ in the primed to the unprimed state is
compatible for the two samples, whereas $\Phi_{\mathrm{0}}$ is
significantly different.

\begin{table}[h!]
\begin{center}
\begin{tabular}{|l|l|l|l|l|}     \hline
name & $\epsilon_{\mathrm{u}} [\%]$, unprimed & $\epsilon_{\mathrm{p}}
[\%]$, primed & $\Phi_{\mathrm{0}}$ [rad] & ratio
$\epsilon_{\mathrm{p}} / \epsilon_{\mathrm{u}}$ \\ \hline \hline A
&$13.7 \pm 0.3$ &$22.3 \pm 0.4$ & $7.4 \pm 0.5$ & $1.63 \pm 0.04 $ \\
\hline B &$11.4 \pm 0.4$ &$18.7 \pm 0.5$ & $108.6 \pm 8.8$ & $1.64 \pm
0.06 $ \\ \hline
\end{tabular}
\end{center}
\caption[]{\label{tab-dose_dep}Summary of the primed and unprimed
collection efficiencies as well as the priming dose $\Phi_0$ and the
ratio of primed to unprimed collection efficiency.}
\end{table}

The primed state of the diamonds has been observed to be stable for
days. The diamonds can be reset to the unprimed state by illumination
with light in the visible region. IR-light was found to have no effect
on the primed state. UV-light with a spectrum extending to about 5~eV
did not show this reset effect but instead a priming effect on the
diamonds.

\begin{figure}[tmb]
\begin{center}
\mbox{\epsfig{figure=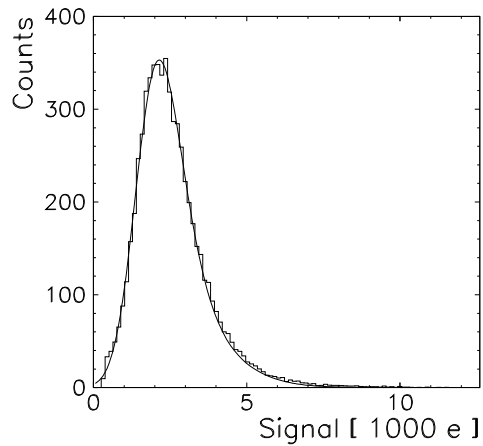,bbllx=0.cm,bblly=0.cm
,bburx=20.cm,bbury=20.cm,width=0.47\textwidth}}
\caption[]{\label{fig-mips_spectrum}Pulse spectrum recorded with sample A.
Superimposed is a smeared Vavilov distribution fitted to the data as
discussed in the text.}
\end{center}
\end{figure}

A pulse-spectrum of sample A in the primed state is shown in
fig.~\ref{fig-mips_spectrum}. By fitting the spectrum with a function,
which is based upon the theoretically expected shape and allows for a
Gaussian distribution of $\epsilon_{\mathrm{Q}}$, information about
the width of $\epsilon_{\mathrm{Q}}$ can be obtained.

A description for the pulse spectrum of MIPs in thin detectors is the
Landau distribution, which does not place a limit on the maximum
energy transfer and thus is only a rough approximation for beta
particles with a maximum energy of $2.28\MeV$ as used in this
experiment. A more general description was derived by
Vavilov~\cite{ref-seltzer}, taking into account the particle velocity,
$\beta$, and the ratio of the mean energy loss of the particle in the
target and the maximum allowed energy transfer $\kappa$ in a single
collision with an atomic electron.  To fit the pulse spectrum, we used
an implementation of this more accurate Vavilov
distribution~\cite{ref-vavilov}, $ V(\eta x'+ \zeta, \kappa, \beta)$,
where $\eta$ is a scaling parameter and $\zeta$ is an offset
parameter. To account for electronic noise and possible intrinsic
variations of $\epsilon_{\mathrm{Q}}$, the Vavilov distribution has
been folded with a Gaussian, $P(\eta(x-x')+\zeta,\eta\sigma)$, with a
width $\sigma$. The complete fit function is given by
\begin{equation}
f(x)= A \int V(\eta x' + \zeta, \kappa, \beta)P(\eta(x'-x)+\zeta
,\eta\sigma)\d x'
\end{equation}
with $A$ being a normalisation factor.  For the fit, we used
$\kappa=0.09$, and $\beta = {v / c} = 0.95$.  The values are motivated
by the assumption that the average energy of the beta particles
contributing to the spectrum was $\bar E_{\beta}=1.7\MeV$.  The
maximum energy transfer was taken to $E_{\mathrm{max}}\approx 1.6\MeV$
since an energy of about 100~keV is required to trigger an event in
the silicon detector. The mean energy loss in the diamond is
approximately given by the specific energy loss of a MIP in diamond.
For a diamond with a thickness of $d=300\mum$, the mean energy loss is
$E_{\mathrm{mean}}\approx140\keV$.  The width of the Gaussian, as well
as the mean $\zeta$, the normalisation $A$ and a scaling parameter
$\eta$ are free variables in the fit.  The fit results in a width of
the Gaussian of $\sigma_{\mathrm{tot}}=490\e$ with $\chi^2 / dof$ of
$1.02$.  To determine the systematic uncertainty of the fit we varied
$0.07 < \kappa < 0.15 $ and $0.86 < \beta < 0.96$ to cover the energy
range for $1.0\MeV < {\bar E}_{\beta} < 2.0\MeV$. This resulted in a
systematic uncertainty on $\sigma_{\mathrm{tot}}$ of
$(^{+30}_{-76})\e$.  The electronic noise was determined to be
$\sigma_{\mathrm{elec.}}=250\e$, thus, the intrinsic broadening of
$\epsilon_{\mathrm{Q}}$ is given by
\[ \sigma_{\mathrm{intr}}=\sqrt{ \sigma_{\mathrm{tot}}^2 - 
\sigma_{\mathrm{elec}}^2}
= (421^{+30}_{-76}) \e \] or expressed in charge collection efficiency
\[ \sigma_{\mathrm{intr}} =  {(421^{+30}_{-76}) \e 
\over \rho_{\mathrm{MIP}} \cdot d} = (4.0^{+0.3}_{-0.7})\% \]
assuming a Gaussian distribution of the broadening.  Translated into
collection distance using equation (\ref{eq-cd}), the width of the
intrinsic Gaussian describing the distribution of the collection
distance is $\sigma_{\mathrm{intr}}= (11.9 ^{+0.8}_{-2.1}) \mum $.
This computation indicates a relative variation of the collection
efficiency, i.e. distance, to its average of $18\%$ at the $1\sigma$
level, which can not be explained by the uncertainty of the fit.  This
is an evidence that $\epsilon_{\mathrm{Q}}$ is not a constant value of
the sample but has a broad distribution.

\section{Measurements with a 10~keV photon beam}
\label{s-hasy-ioni}
The knowledge of the local variation of $\epsilon_{\mathrm{Q}}$ for
homogeneous ionisation is important for the application as a position
sensitive detector for highly energetic particles, where strong
inhomogeneities would deteriorate the efficiency and position
resolution of the device.  In the following, photons are used to
measure the average charge collection properties of diamond samples.
We present a measurement of $\epsilon_{\mathrm{Q}}$ which is spatially
resolved over the active detector area. This allows a study of the
homogeneity of $\epsilon_{\mathrm{Q}}$ on a scale of about $100\mum$.

Photons of 10 keV interact with diamond predominantly via the
photo-effect. When averaged over many photons the ionisation density
distribution is given by $\rho(x)=A_0 e^{-{x \over \lambda_0}}$ with
$\lambda_0$ being the absorption length, $x$ the distance in the
material, and $A_0$ a constant. Since $\lambda_0=1300\mum$ for 10 keV
photons in diamond and the thickness is $d=300\mum$ for the
investigated sample A, the resulting ionisation density distribution
is approximately homogeneous.

\subsection{Experimental set-up}
A 10 keV photon beam from the HASYLAB synchrotron light source at DESY
was used to study the charge collection properties of CVD diamond.
Monochromatic photons with an energy of $E=(10\pm0.5)\keV$ were
selected using a graphite monochromator. The beam size was reduced
with a variable aperture to a spot size of typically $100\mum \times
100\mum$.

\begin{figure}[tmb]
\begin{center}
\mbox{\epsfig{file=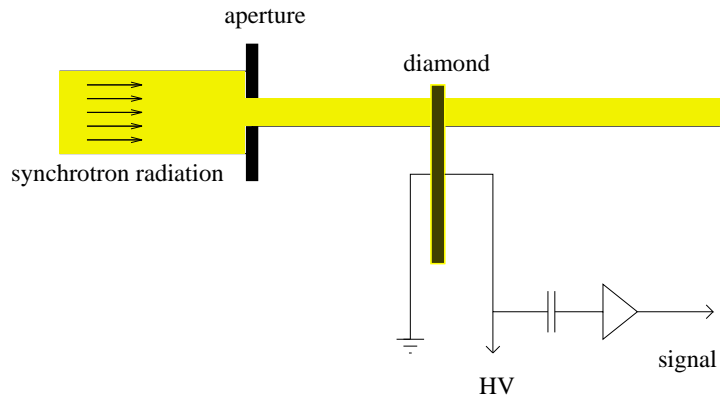,bbllx=0cm,bblly=0.cm
,bburx=481pt,bbury=261pt,width=0.7\textwidth}}
\caption[]{\label{fig-hasy-ioni-setup}Schematic view of the set-up for
the measurement with synchrotron radiation.}
\end{center}
\end{figure}

The beam clock of the synchrotron provides an external trigger source
marking the arrival of a photon pulse every 500 ns.  The duration of
the photon pulses is in the order of picoseconds.  The flux per bunch
of the monochromatic synchrotron beam is calculated to be of the order
of $10^4$ ${\gamma \mm^{-2}}$.  The expected charge deposition in the
diamond is $0.5\pC\mm^{-2}$ per bunch.  The diamond was mounted on a
movable table behind the aperture.  An electrical field was applied to
the diamond and the photoinduced signals were amplified and read out
with the same read-out chain as described above, except that the
shaping time was reduced to 500 ns with a shaping amplifier (Ortec
454).  The chain was calibrated by injecting a known charge using a
test-pulser and a capacitor. The signals of the amplifier chain were
averaged over 2048 bunches to improve the signal-to-noise ratio.  A
schematic view of the set-up is shown in
fig.~\ref{fig-hasy-ioni-setup}.

\begin{figure}[tmb]
\begin{center}
\setlength{\unitlength}{1cm}
\begin{picture}(7.88,5.344)
\put(0,0){ \mbox{\epsfig{file=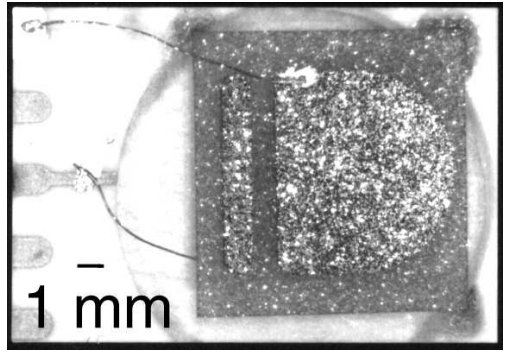,bbllx=36pt,bblly=207pt
,bburx=575pt,bbury=585pt,width=0.47\textwidth}} }
\end{picture}
\caption[]{\label{fig-pattern}The investigated sample A with its
metallisation pattern, mounted on a ceramic holder.}
\end{center}
\end{figure}

The CVD diamond sample A was measured in this set-up. It was
metallised on both sides with a pattern as shown in
fig~\ref{fig-pattern}.

\subsection{Results and discussion}
\begin{figure}[tmb]
 \begin{center}
  \begin{tabular}{cc}

\subfigure[ ] {\epsfig{figure=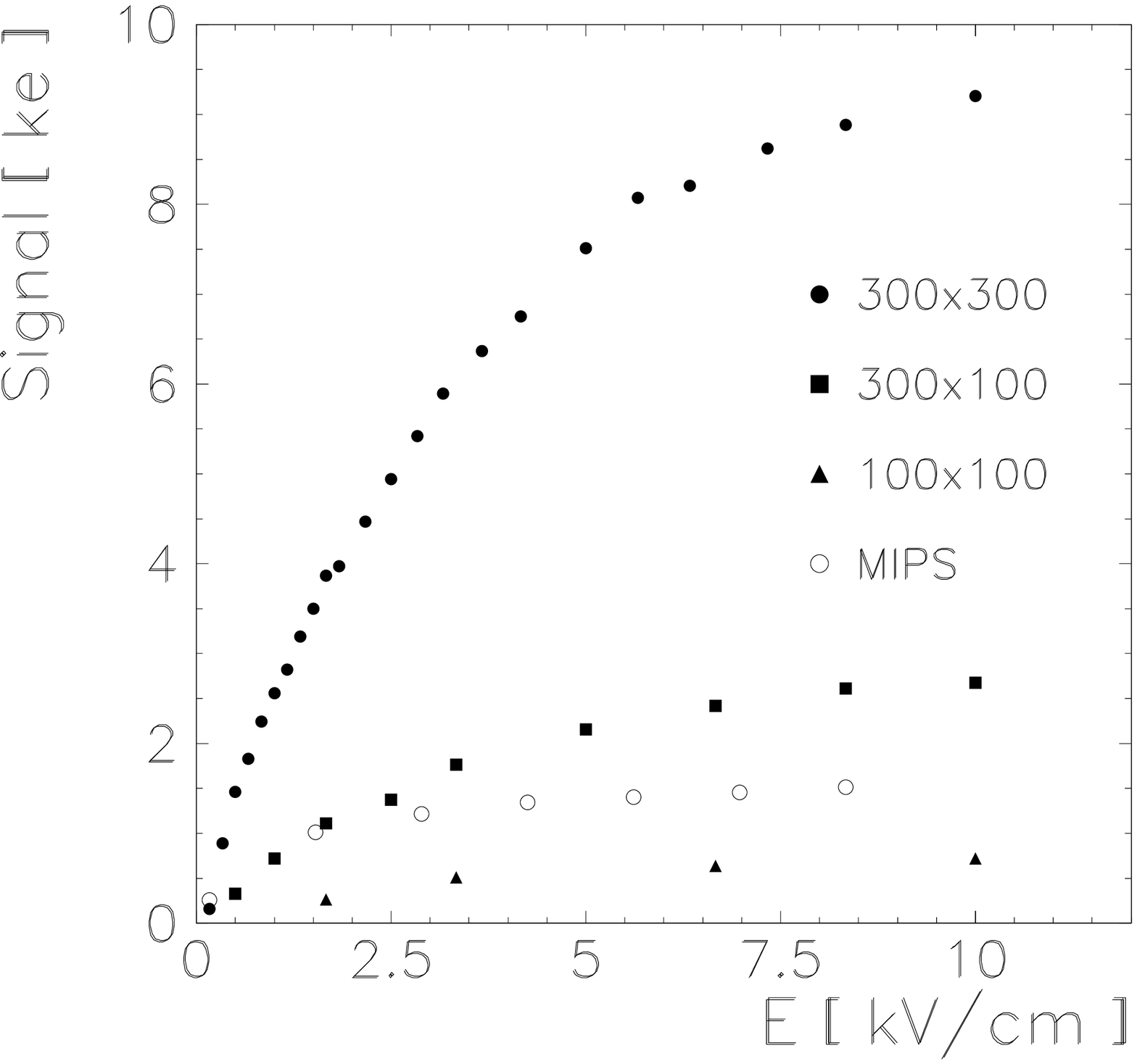,bbllx=0.cm,bblly=0.cm
,bburx=20cm,bbury=20cm,width=0.47\textwidth}}

\subfigure[ ] {\epsfig{figure=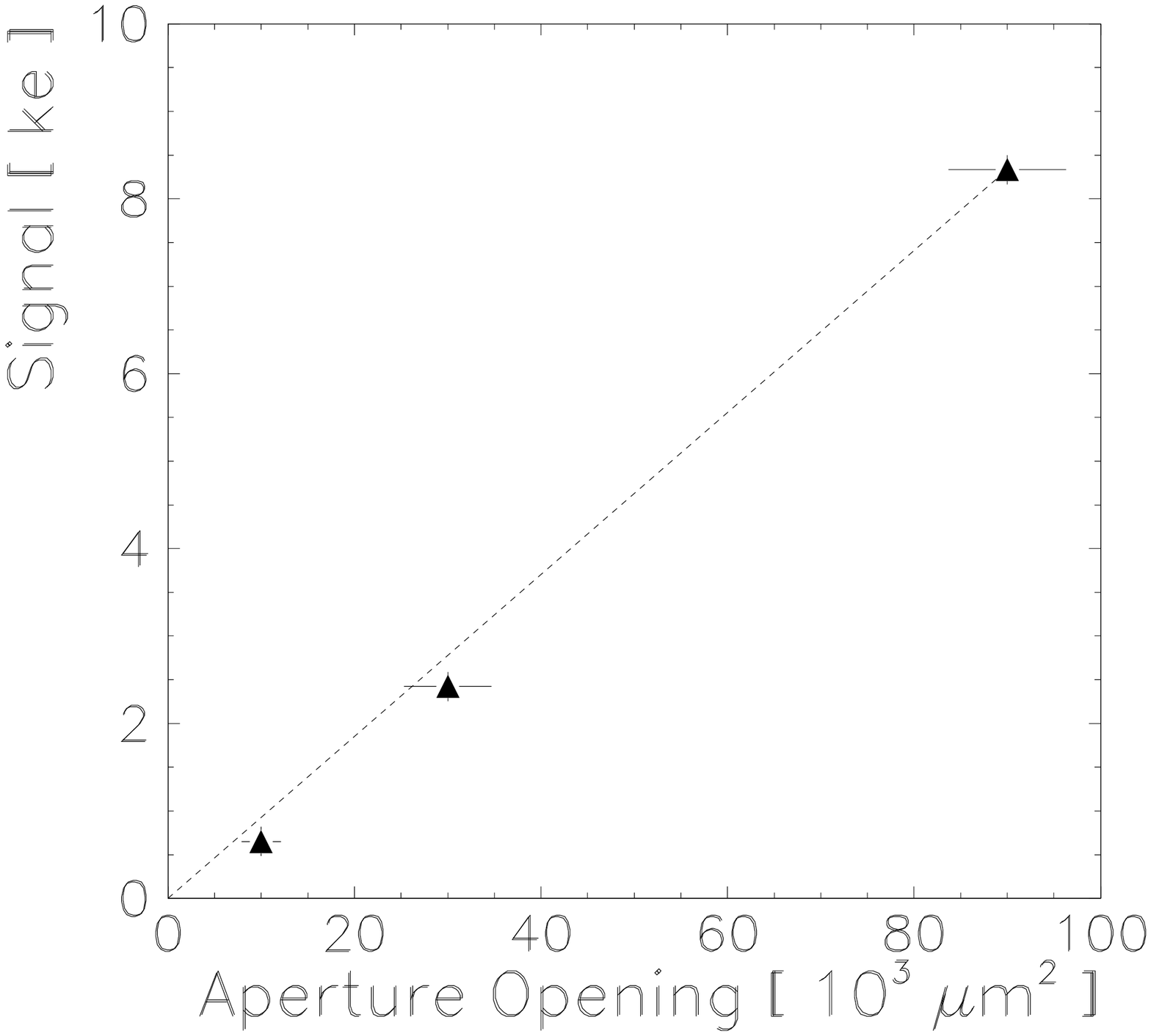,bbllx=0.cm,bblly=0.cm
,bburx=20cm,bbury=20cm,width=0.47\textwidth}}

  \end{tabular}
 \end{center}
\caption[]{\label{fig-hasyhasy}(a) Photoinduced signal of the 
diamond detector
vs electrical field strength for different aperture openings. For
comparison, the signal obtained with MIPs is depicted too. (b) Signal
for $E=10\kV/ \cm$ as a function of the aperture opening.  The dashed
line indicates the expected progression for a linear behaviour. }
\end{figure}

We have measured the response signal at different field strengths and
for different aperture openings.  Three different aperture sizes of
approximately $300\mum \times 300\mum$, $300\mum \times 100\mum$ and
$100\mum \times 100\mum$ were used.  Fig.~\ref{fig-hasyhasy}a shows
the induced signal as a function of the applied field. The dependence
is very similar to the results obtained with MIPs (see
section~\ref{s-mips}).  A steep rise of the signal at small values of
$E$ is followed by a flattening of the curve at higher field
strengths.

Within the precision of the aperture adjustment, a linear scaling of
the induced signal with the spot size has been observed
(fig.~\ref{fig-hasyhasy}b).  Furthermore, the signal height is
compatible with the magnitude calculated from the photon flux and the
charge collection efficiency measured with MIPs (see
section~\ref{s-mips}).  Although no precise absolute values of
$\epsilon_{\mathrm{Q}}$ can be determined due to the uncertainties in
the absolute photon flux, accurate measurements of relative changes of
$\epsilon_{\mathrm{Q}}$ across the surface of the detector are
obtained. Since the ionisation density distribution is similar to the
distribution created by MIPs, we assume a similar charge collection
behaviour with 10 keV photons. Therefore, we normalise the mean of the
measured distribution of $\epsilon_{\mathrm{Q}}$ to the value measured
with beta particles and can calibrate this way the set-up.

To obtain a spatial map of $\epsilon_{\mathrm{Q}}$, we scanned the
metallised part of the sample A, consisting of a half-circle and
rectangle (see fig.~\ref{fig-pattern}), with a spot size of $100\mum
\times 100\mum$ and with a step size of $100\mum$. At each point, the
induced signal for a field of $9\kV/\cm$ has been measured.  The
signal has been corrected for the change in photon flux in time,
deduced from the beam current in the electron storage ring.

\begin{figure}[tmb]
 \begin{center}
  \begin{tabular}{cc}
 \subfigure[ ] {\epsfig{figure=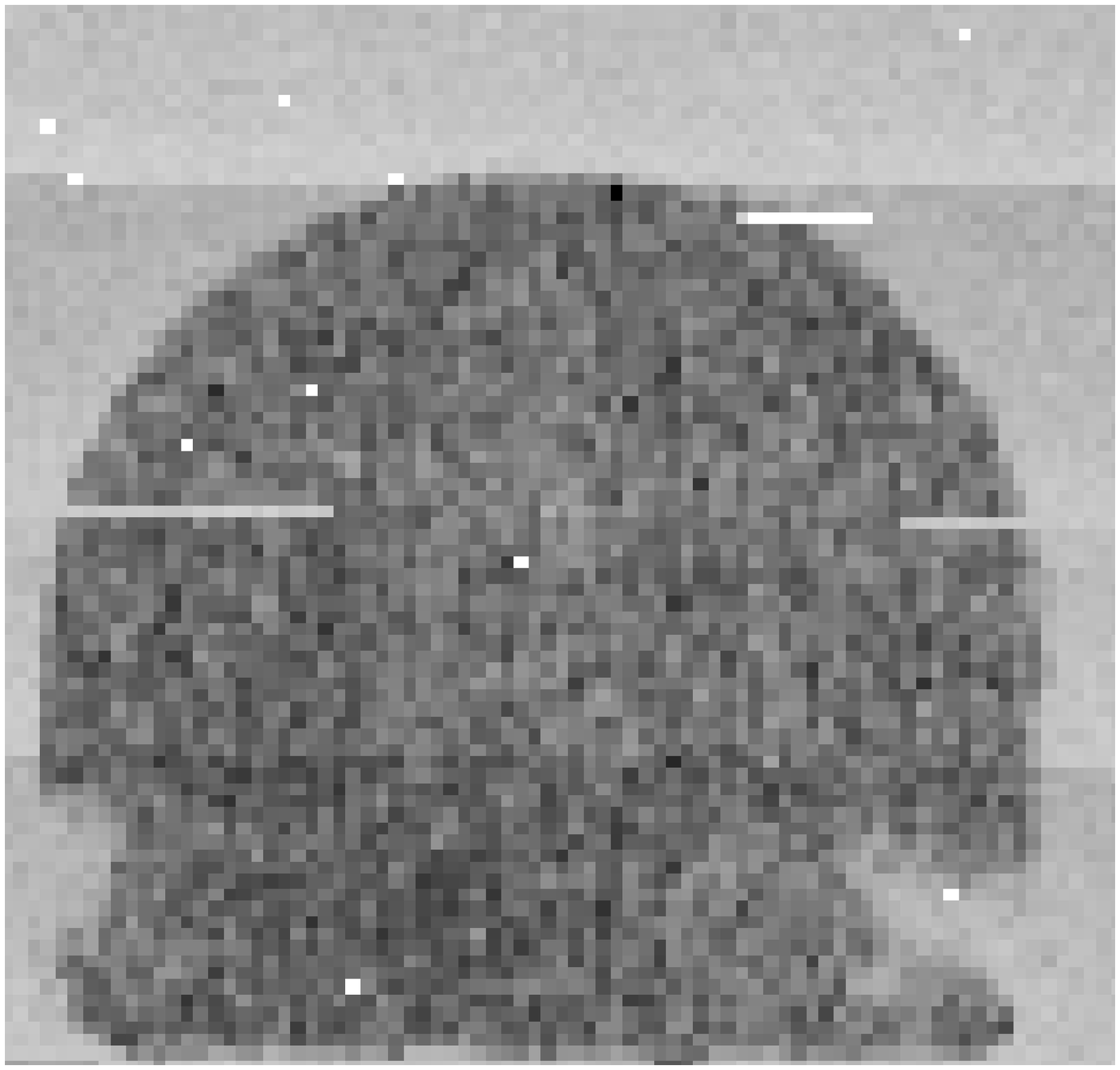,bbllx=36pt,bblly=124pt
,bburx=574pt,bbury=666pt,width=0.35\textwidth}} \subfigure[ ] {
\setlength{\unitlength}{1cm}
\begin{picture}(8,6)
\put(0,0){ \epsfig{figure=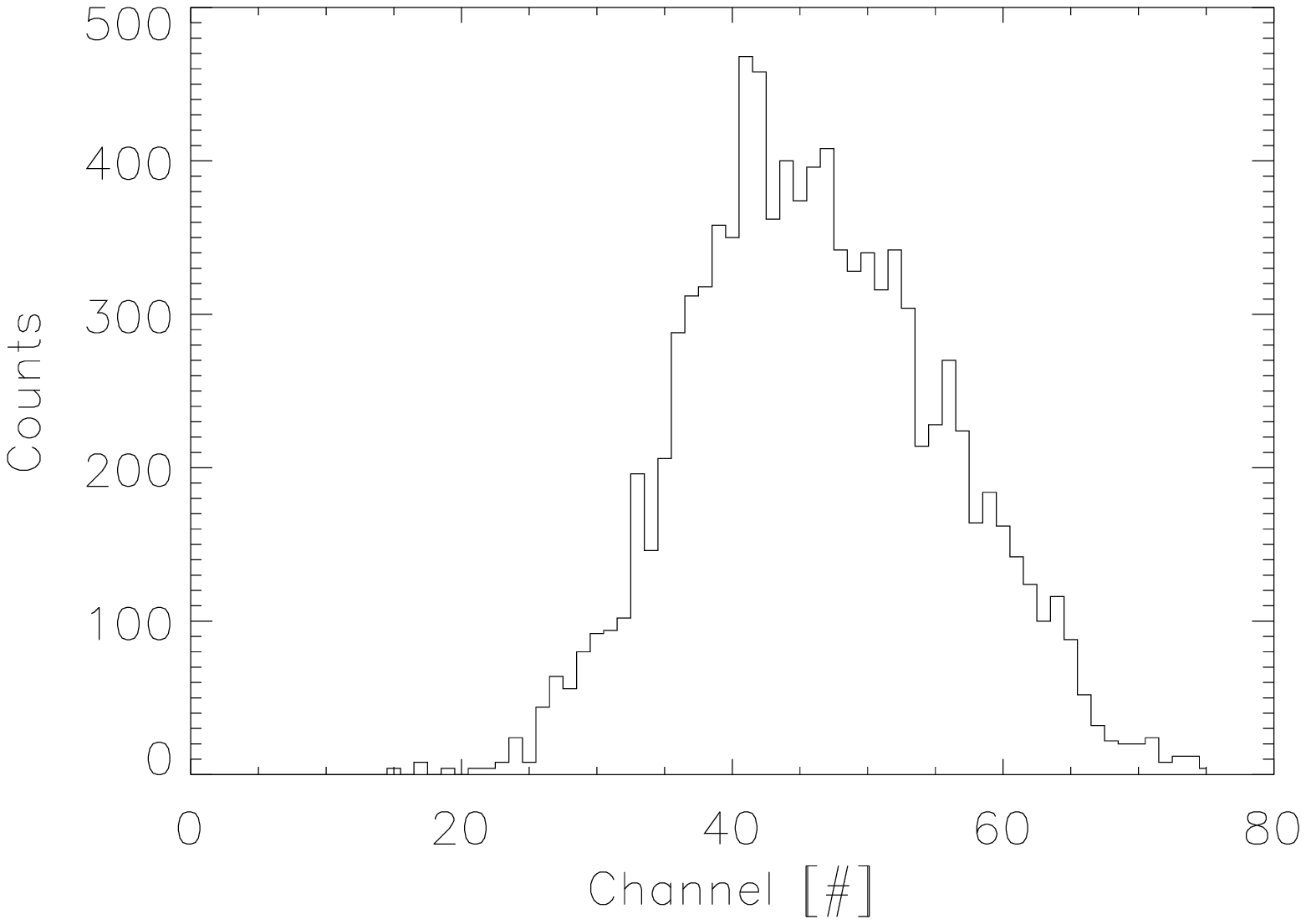,bbllx=89pt,bblly=371pt
,bburx=543pt,bbury=696pt,width=0.65\textwidth} }
\end{picture}
}
  \end{tabular}
 \end{center}
\caption[]{\label{fig-hanswurst} (a) Spatially resolved 
representation of the
 photoinduced signal. The picture represents a region of $8\mm \times
8\mm$.  The white spots and stripes are measurement failures of the
DAQ system.  (b) Distribution of the photoinduced response signal from
the metallised region of the diamond detector.}
\end{figure}

The map obtained is displayed in fig.~\ref{fig-hanswurst}a.  The
signal heights are encoded using a grey-scale.  High signals are
represented by dark grey, areas of low signals are marked with a light
grey.  The metallised part of the sample gives a clear signal, while
the non-metallised area show only the pedestal noise. The light areas
on the lower left and right side are caused by absorption from silver
paint, which has been used to contact the sample.  The spectrum of the
signal response on the metallised area is shown in
fig.~\ref{fig-hanswurst}b. It represents the distribution of the
measured $\epsilon_{\mathrm{Q}}$ on the metallisation folded with a
distribution, which is due to noise.  As the uncertainties in the
photon flux allow only relative measurements, we follow the discussion
above and set the average of the distribution equal to
$\epsilon_{\mathrm{Q}}=(22.3\pm 0.4)\%$ measured with MIPs (see
sec.~\ref{s-mips}) assuming that the sample was in a primed state. The
expected dose absorption of the diamond sample is of the order of
$1\krad \s^{-1}$ and thus the sample should be fully primed in less
than 0.1 second as the priming dose constant was measured to
$\Phi_0=7.4\pm0.5\rad$ (see sec.~\ref{s-mips}).  To account for the
uncertainty of the priming state of the sample due to the exposure to
the beam we rescanned the diamond a second time and found a shift in
the mean of the spectrum of factor 1.05 which is assigned as a
systematic error.

The noise distribution has been estimated from the data by looking at
the spread of the amplitude for 8 subsamples. The average of the
spread is $\sigma_{\mathrm{noise}}=(0.67\pm 0.03)\%$.  To determine
the total width of the signal, a Gaussian has been fitted to the
distribution. A width of $\sigma_{\mathrm{m}}=(4.37 \pm 0.21)\%$ is
obtained from the fit. To extract the intrinsic width of the
$\epsilon_{\mathrm{Q}}$ distribution, we assumed the contributions of
the noise and the intrinsic variations of $\epsilon_{\mathrm{Q}}$ to
be Gaussian and calculated the intrinsic width of the
$\epsilon_{\mathrm{Q}}$ distribution by
\[\sigma_{\mathrm{intr}} = \sqrt{\sigma_{\mathrm{m}}^2 - 
( {\sigma_{\mathrm{noise}} \over \sqrt{8}}) ^2}  = (4.32 \pm 0.22)\%.\]

This value has to be compared with a measurement of
$\sigma_{\mathrm{intr}}$ with MIPs as discussed in
section~\ref{s-mips} which gives a value of
$\sigma_{\mathrm{intr}}=(4.0^{+0.3}_{-0.7})\% $.  The two methods give
values of $\sigma_{\mathrm{intr}}$ which are in good agreement.  The
high value of $\sigma_{\mathrm{intr}}$ reflects the polycrystalline
nature of the CVD diamond.

As the measurement with beta-particles were performed with an aperture
of $1\mm$ the variation of $\sigma_{\mathrm{intr}}$ was calculated as
a function of the position of a $1\mm \times 1\mm$ box, which
contained the events used for the calculation of
$\sigma_{\mathrm{intr, box}}$. Within the statistical error no
deviation from the result calculated with full statistics was
observed.

\section{Measurements with alpha particles}
\label{s-alpha}
The crystallites in polycrystalline CVD diamond films are larger on
the growth side than on the substrate side. Therefore, it is
interesting to study the charge collection properties on both sides
separately. This was done with mono-energetic alpha particles from a
\Cm source with an energy of $5.8\MeV$. The mean free path of
$5.8\MeV$ alpha particles in diamond is about $16\mum$.  Most of the
kinetic energy is deposited via ionisation at the end of the track
(Bragg-peak).  The deposition of charge takes place in a shallow layer
close to the surface, whose charge collection properties determine the
signal.  Thus, the response signal to alpha particles can be used to
extract information about the difference of $\epsilon_{\mathrm{Q}}$ on
growth and substrate side.

\subsection{Experimental set-up}
The alpha-source and the diamond were placed in a vacuum-chamber at a
pressure of $10^{-5}\mbar$.  The AC-coupled amplifier chain was
connected to the side which has been exposed to the alpha-particles.
The polarity of the electrical field is defined relative to the
exposed side of the diamond, i.e.~a negative potential on the exposed
side is a negative field.  The detector signal was amplified by a
Canberra 2022 charge sensitive preamplifier, followed by an Ortec 572
shaping-amplifier with a time constant of $\tau=3\mus$ and an Ortec
444 gated biased amplifier. Since the mean free path of the alpha
particles is much smaller than the film thickness of the diamond no
external trigger could be used.  The signal from the diamond itself
had to be used as a trigger, which required to set a threshold to
discriminate events from noise. The threshold was set to 6700 e. The
noise from the system was determined to be 590 e.  The signals were
recorded with a multi-channel-analyser. A silicon detector was used to
calibrate the set-up, assuming 100\% collection efficiency of the
silicon detector and an energy of $3.6\eV$ to create an
electron-hole-pair at room temperature in silicon.

\subsection{Experimental results}
\begin{figure}[tmb]
 \begin{center}
  \begin{tabular}{cc}

\subfigure[ ] {\epsfig{figure=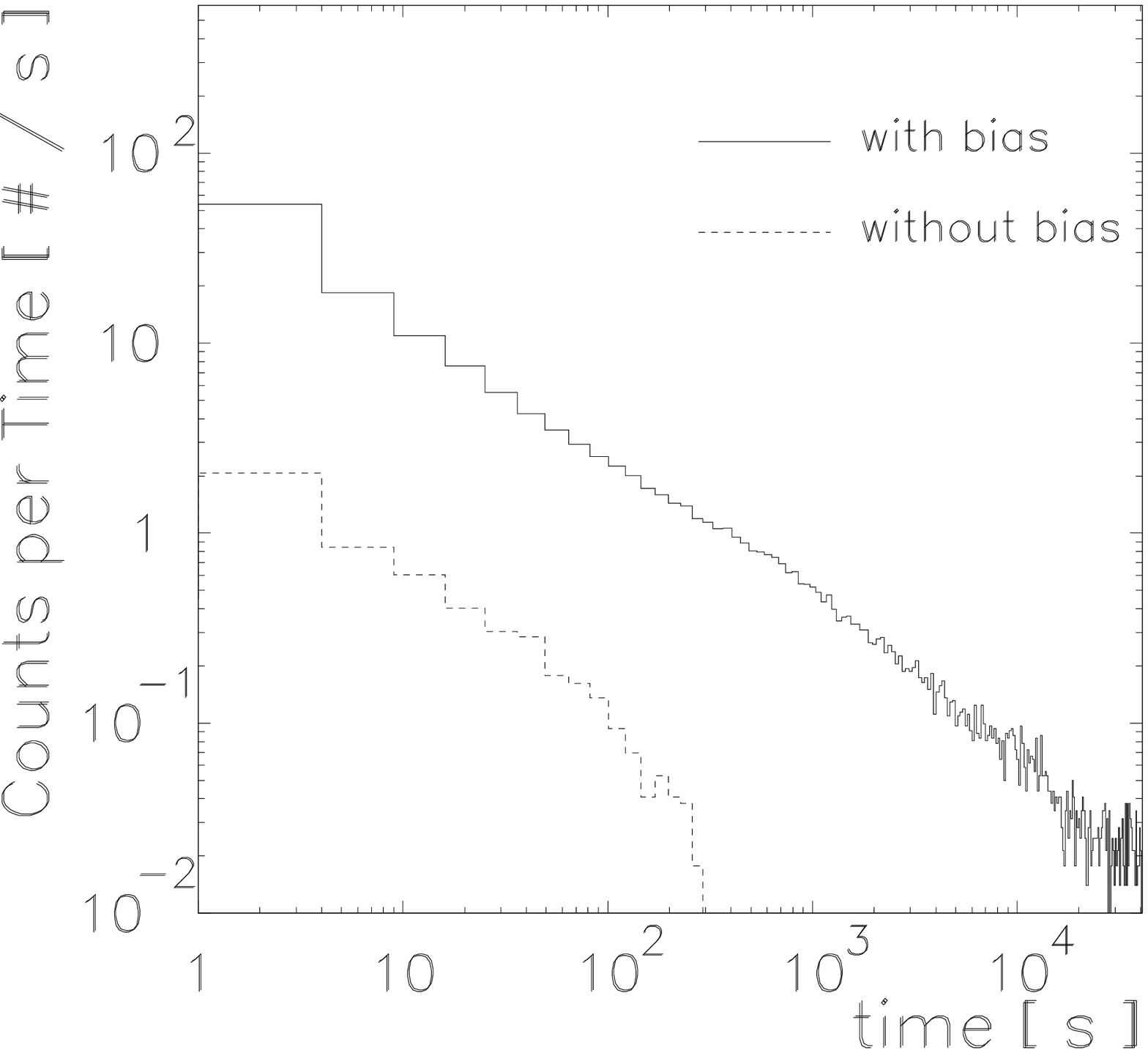,bbllx=0.cm,bblly=0.cm
,bburx=20cm,bbury=20cm,width=0.47\textwidth}}

\subfigure[ ] {\epsfig{figure=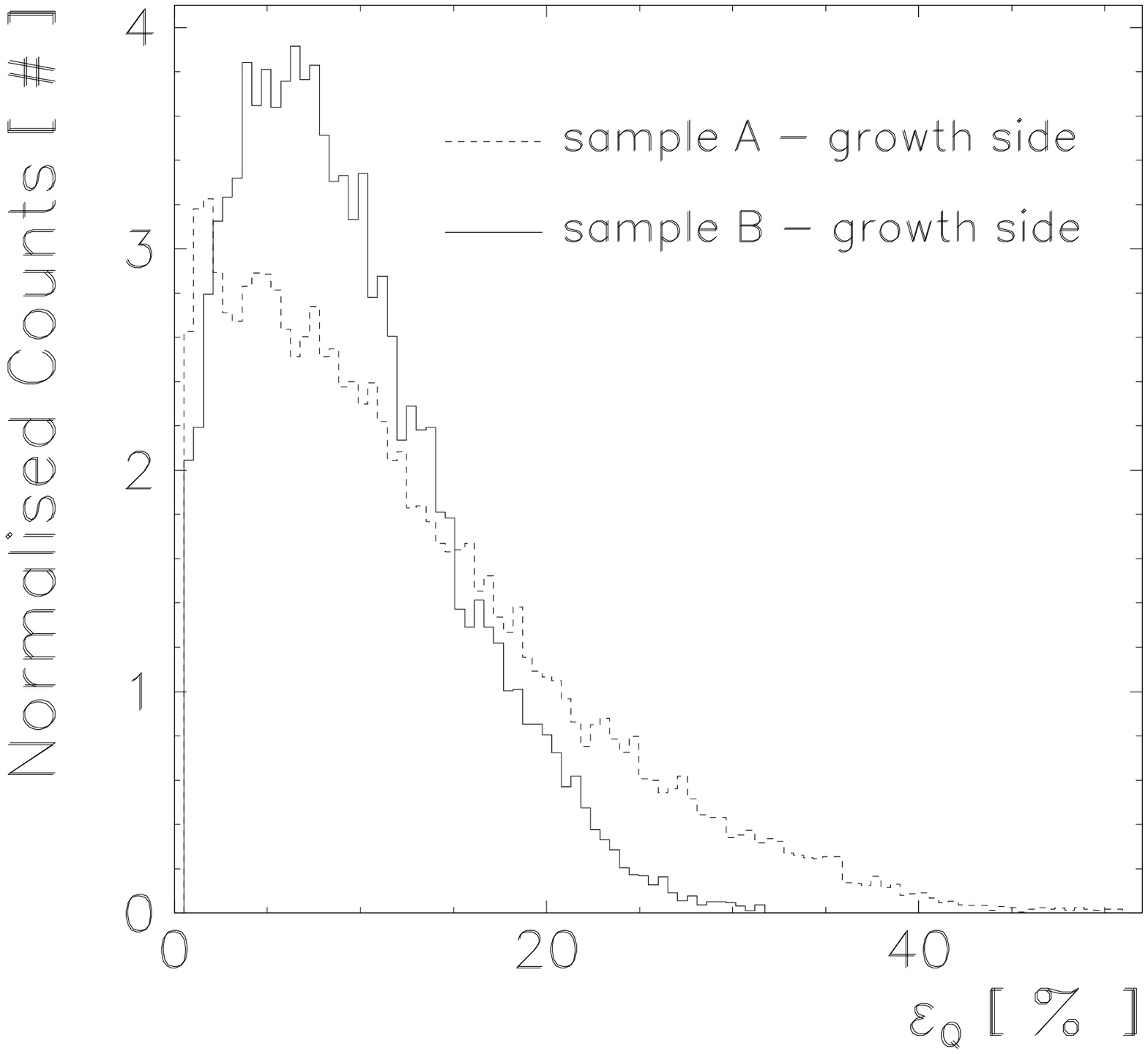,bbllx=0.cm,bblly=0.cm
,bburx=20cm,bbury=20cm,width=0.47\textwidth}}

  \end{tabular}
 \end{center}
\caption[]{\label{fig-bla}(a) Counting rate as a function of time
for applied voltage ($U_{\mathrm{b}}=250\V$) and polarisation field
($U_{\mathrm{b}}=0\V$), measured with sample A. (b) Typical
alpha-spectra obtained with CVD diamond.}
\end{figure}

Alpha-spectra for different electrical field strengths have been
recorded, exposing the growth as well as the substrate side.  During
the irradiation with alpha particles, the counting rate and the pulse
height of the diamond detectors were decreasing. The change of the
counting rate with time for a fixed threshold can be seen in
fig.~\ref{fig-bla}a.  The decreasing counting rate indicates the
build-up of a polarisation field which compensates the externally
applied field, as after removing the voltage, the diamond was still
polarised and signals of opposite polarity have been observed.

Since the polarisation lowers the signal over the exposure time, the
measured $\epsilon_{\mathrm{Q}}$ depends on the history of the sample.
To have reproducible results, the diamond was exposed for a period of
$300\s$ to alpha particles while having applied the electrical
field. Then the field was reversed and the response for the same time
was measured.  The counting rate dropped more than an order of
magnitude during the measurement period.  This procedure ensured a
well-defined initial condition of the sample, independent of the
initial polarisation state.

The two samples, A and B, show a different response. The $300\mum$
thick sample A exhibits a steeply falling spectrum with no visible
peak when irradiated with mono-energetic alpha particles, while the
$123\mum$ thick sample B produces a very broad peak.  Typical spectra
are shown in fig.~\ref{fig-bla}b. As already pointed out in
\cite{ref-manfredotti96}, the interpretation of spectra in terms of
$\epsilon_{\mathrm{Q}}$ is difficult due to the influence of
polarisation, which is disturbing the intrinsic electronic properties
of the crystallites. The method is especially sensitive to
polarisation because of the thin layer in which most of the charge is
deposited. The measurements of the $\epsilon_{\mathrm{Q}}$
distribution using methods which are less sensitive to polarisation
(beta particles, 10~keV photons) show a much narrower distribution of
$\epsilon_{\mathrm{Q}}$.

Nevertheless, the data can be used to compare the response of growth
and substrate side at different field strengths and polarities.  To
extract a value of $\epsilon_{\mathrm{Q}}$ from the alpha spectra we
use two methods.  For all spectra, we take the average $\bar x = [\int
x \cdot f(x) \d x] / [\int f(x) \d x]$ of the spectrum as a measure of
$\epsilon_{\mathrm{Q}}$. Since we apply a noise cut in the spectrum,
the observed average $\epsilon_{\mathrm{Q}}$ for alpha particles
should be systematically higher than the true $\epsilon_{\mathrm{Q}}$.
If a peak is visible, we additionally fit a Gaussian to the spectrum
and take the mean as another measure of $\epsilon_{\mathrm{Q}}$.  This
can only be done for sample B.

\begin{figure}[p]
 \begin{center}
  \begin{tabular}{cc}
 \subfigure[ ] {\epsfig{figure=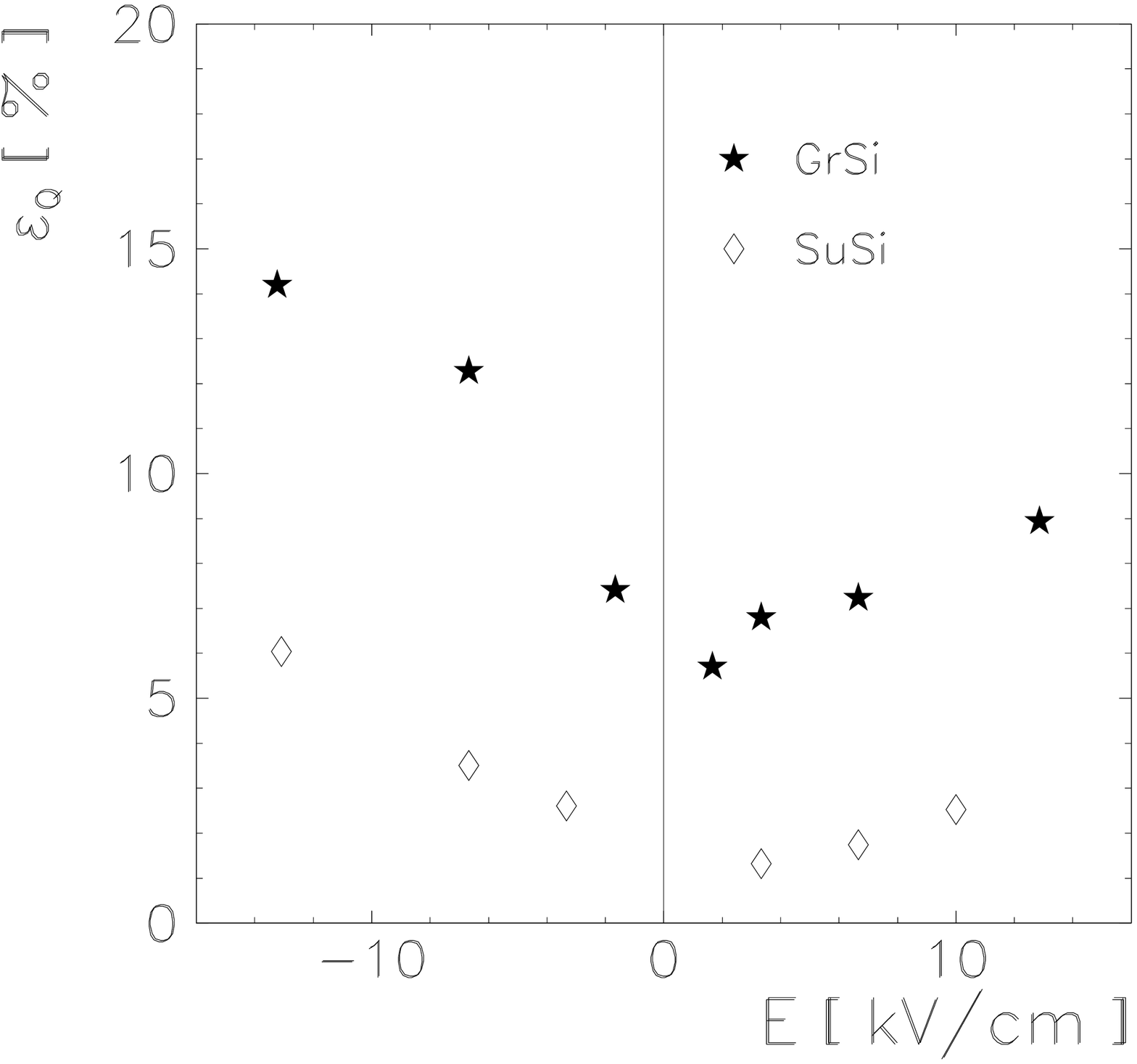,bbllx=0.cm,bblly=0.cm
,bburx=20.cm,bbury=20.cm,width=0.47\textwidth}} & \subfigure[ ]
{\epsfig{figure=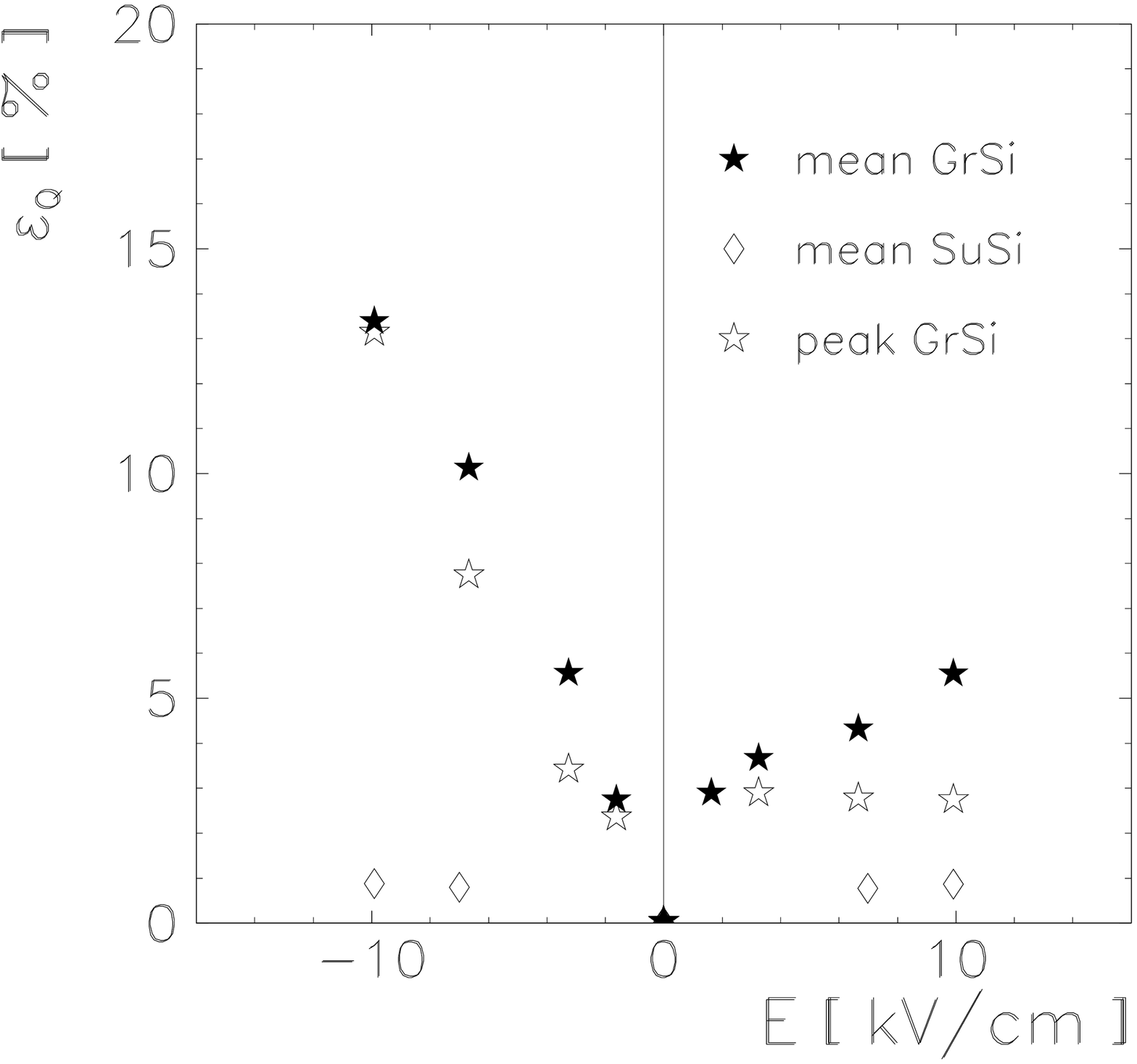,bbllx=0.cm,bblly=0.cm
,bburx=20.cm,bbury=20.cm,width=0.47\textwidth}}
  \end{tabular}
 \end{center}
\caption[]{\label{fig-analyze}Field dependence of the collection
efficiency measured at the substrate (SuSi) and growth sides (GrSi)
for sample A (a) and B (b), determined with alpha particles by taking
the mean of the spectrum. In the case of sample B, a Gaussian fit was
performed in addition if a peak was visible.}
\end{figure}

The field strength dependence of $\epsilon_{\mathrm{Q}}$ is shown for
both samples in fig.~\ref{fig-analyze}.
One should note that the dependence is similar to the measurements
with beta particles.  Furthermore, $\epsilon_{\mathrm{Q}}$ is higher
for negative fields than for positive fields on both growth and
substrate side indicating a higher mobility lifetime product for
electrons.

The response is much better on the growth side than on the substrate
side, as expected from the better crystalline quality on the growth
side. The results from growth and substrate side can be compared to
verify the linear model proposed in~\cite{ref-zhao}, which predicts a
linear increase of the collection distance with the film thickness. If
we apply this model to sample A, the linear model predicts a ratio of
growth side to substrate side collection distance of $R={500\mum /
200\mum}\approx 2.5$, taking into account that sample A was ground at
the substrate side from an original thickness of $500\mum$ to
$300\mum$.  The measured values are varying between 2.4 at $E=13.3{\kV
/ \cm}$ and 5.2 at $E=3.3{\kV / \cm}$, with a mean value of
$R_{\mathrm{m}} = 3.7 \pm 0.4$.

The average $\epsilon_{\mathrm{Q}}$ for negative and positive polarity
at a field of $E=10{\keV / \cm}$ is $<\epsilon_{\mathrm{Q}}>=10.1\%$
for sample B and $<\epsilon_{\mathrm{Q}}>=7.1\%$ for sample A. Using
the van~Hecht-equation~(\ref{eq-hecht}), the average collection
distance for sample A is approximately $\delta=43\mum$ and for sample
B $\delta=25\mum$. Comparing these values with measurements of
$\delta$ obtained with beta particles, the agreement is reasonable for
A and poor for B.  Beta particles enable a more precise measurement of
$\delta$ since polarisation effects are negligible.

\section{Measurements with a proton micro-beam}
\label{s-pixe}
A proton micro-beam~\cite{ref-blabla} of $2\MeV$ protons was used to
investigate the lateral distribution of $\epsilon_{\mathrm{Q}}$ on the
growth and substrate side.  The penetration depth of $2\MeV$ protons
in diamond is about $30\mum$~\cite{ref-zieglerp} and, therefore, much
smaller than the sample thickness ($d=300\mum$).  The ionisation
density has a characteristic Bragg peak at the end of the trajectory.
Only a shallow layer of the diamond of a few microns thickness is
probed, similar to the investigations using alpha-particles. Thus the
information about the $\epsilon_{\mathrm{Q}}$ distribution and
homogeneity is disturbed by a strong polarisation field. The spatially
resolved response behaviour of growth and substrate side was recorded
and compared to SEM pictures. Only sample A was investigated.

\subsection{Experimental set-up}
The proton micro-beam had a proton energy of $2\MeV$ and a beam spot
size of about $1\mum$. The sample has been mounted on a sample-holder
in an evacuated chamber at a pressure of about $10^{-5}\mbar$.

The diamond signal was amplified by an Amptek A250 charge integrating
amplifier. Then a shaping amplifier ($\tau=1\mus$) and a gated biased
amplifier, also used for the alpha particle measurements, further
amplified the signal. The signal was again self-triggered with the
threshold set to $2400\e$. The set-up has been calibrated with a
silicon-detector.

\subsection{Experimental results}
\begin{figure}[tmb]
\begin{center}
\mbox{\epsfig{figure=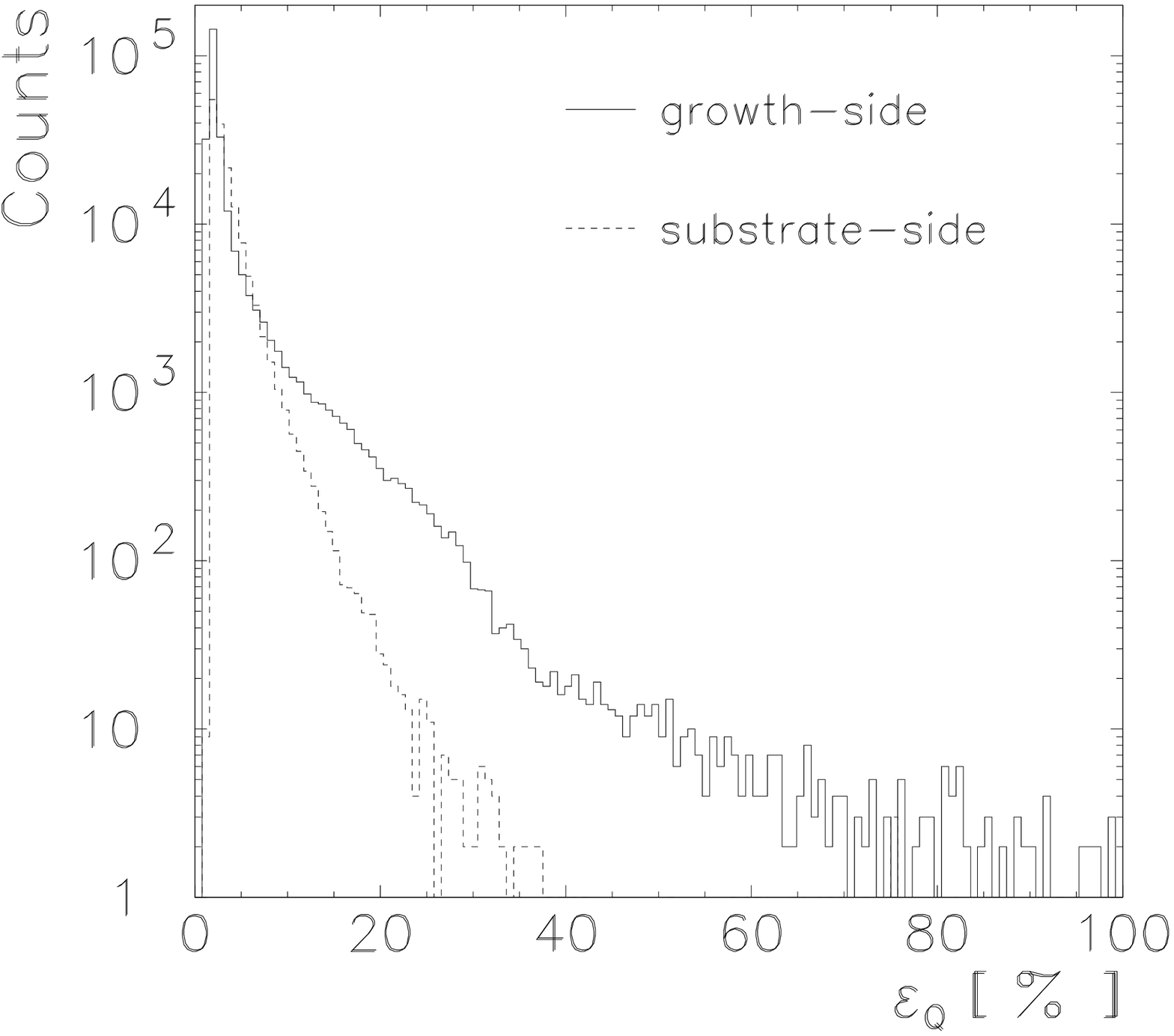,bbllx=0.cm,bblly=0.cm
,bburx=20cm,bbury=20cm,width=0.47\textwidth}}
\caption[]{\label{fig-spec-growth-subs}Response spectra of
       sample A to 2 MeV protons at $E=5.7{\kV / \cm}$ for growth and
       substrate side.}
\end{center}
\end{figure}

The response spectrum of the diamond film to $2\MeV$ protons at
$E=5.7{\kV / \cm}$ is displayed in fig.~\ref{fig-spec-growth-subs} for
the substrate and the growth side.

The shape of the spectra is similar to those seen with alpha
particles.  Furthermore, the substrate side is less efficient than the
growth side.  The ratio of the average pulse heights of growth and
substrate side is $R=1.8$, which is smaller than the ratio observed
with alpha particles (see section~\ref{s-alpha}). The difference could
be due to the lower threshold and the bigger penetration depth of
$2\MeV$ protons compared to $5.8\MeV$ alpha particles.

Unlike the alpha measurements, measurements with the proton micro-beam
allow to resolve signals spatially thus obtaining a map of
$\epsilon_{\mathrm{Q}}$.

The edges of the metallisation were clearly visible and some spot-like
structures of high response with typical dimensions of $50\mum$ to
$100\mum$ with a spot density of about $1\mm^{-2}$ were observed.  The
vast majority of the sample did not show a signal above the applied
threshold of the read-out system.  The fact that the regions at the
edges of the metallisation gave a higher response is attributed to the
higher field strength due to the non-uniformity of the electric
field. We made use of this effect to correlate proton micro-beam
measurements with SEM pictures, as discussed below.

The centres of high response, which are named in the following {\em
hot spots}, were partly not stable in time, indicating polarisation
phenomena. Some remained localised and constant under continuous
irradiation, others tended to disappear, while yet another type seemed
to appear only after some irradiation. Measurements with a proton
micro-beam from \cite{ref-manfredotti96} are in agreement with our
observations.

\begin{figure}[tmb]
\begin{center}
\setlength{\unitlength}{1cm}
\begin{picture}(6,6.7)
\put(-1,0){ \epsfig{file=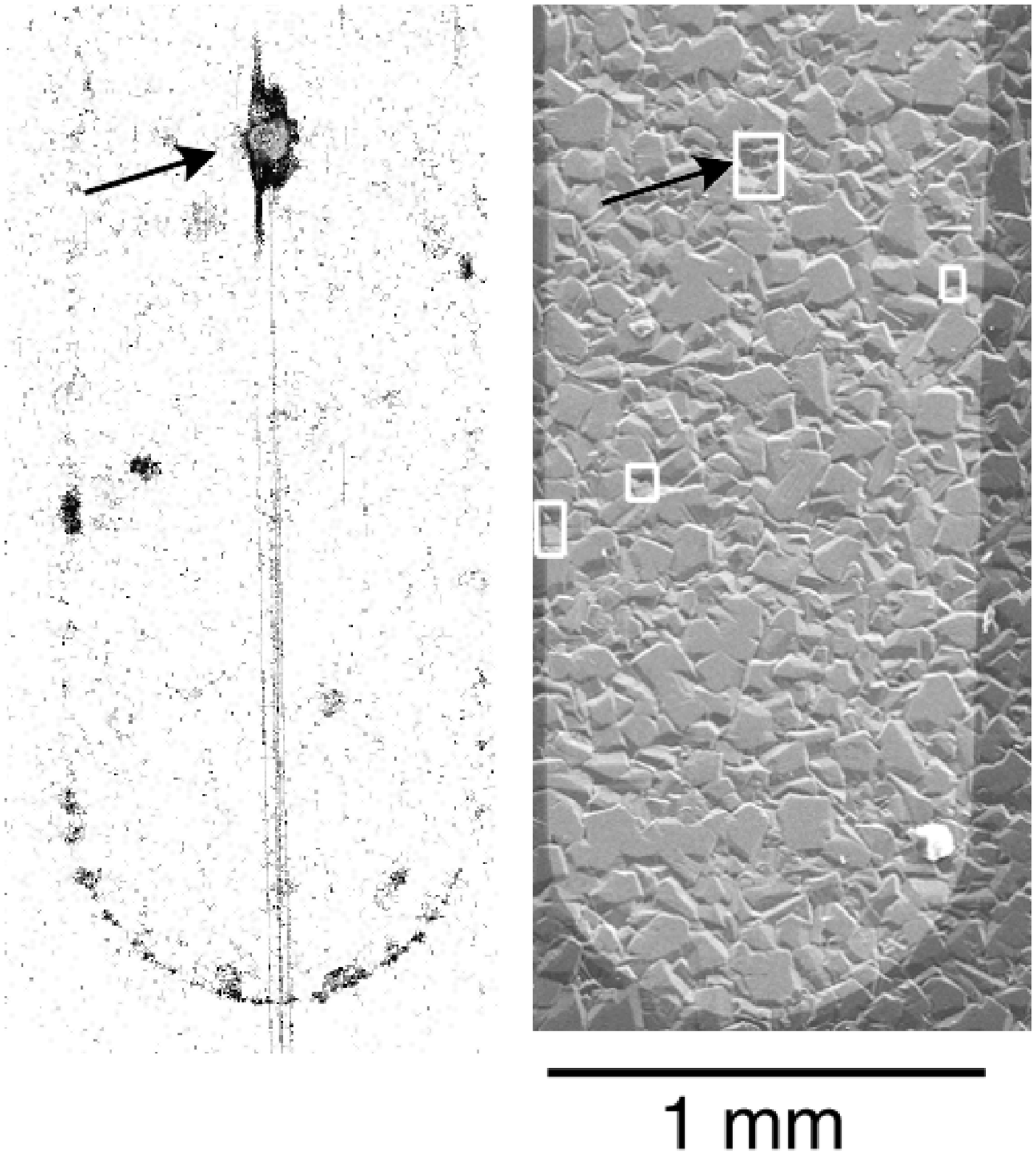
,bbllx=32pt,bblly=127pt,bburx=555pt,bbury=664pt,width=0.55\textwidth}
}
\end{picture}
\caption[]{\label{fig-pix-sem}Correlation of the map
obtained with protons (left) with SEM picture (right).  Dark areas
correspond to a high response (hot spots).  Horizontal stripes are
artifacts of the measurement.  The framed areas on the SEM mark the
locations of the hot spots. The arrow points to the hot spot $X$
discussed in the text. }
\end{center}
\end{figure}

\begin{figure}[tmb]
 \begin{center}
  \begin{tabular}{cc}
 \subfigure[ ] {\epsfig{figure=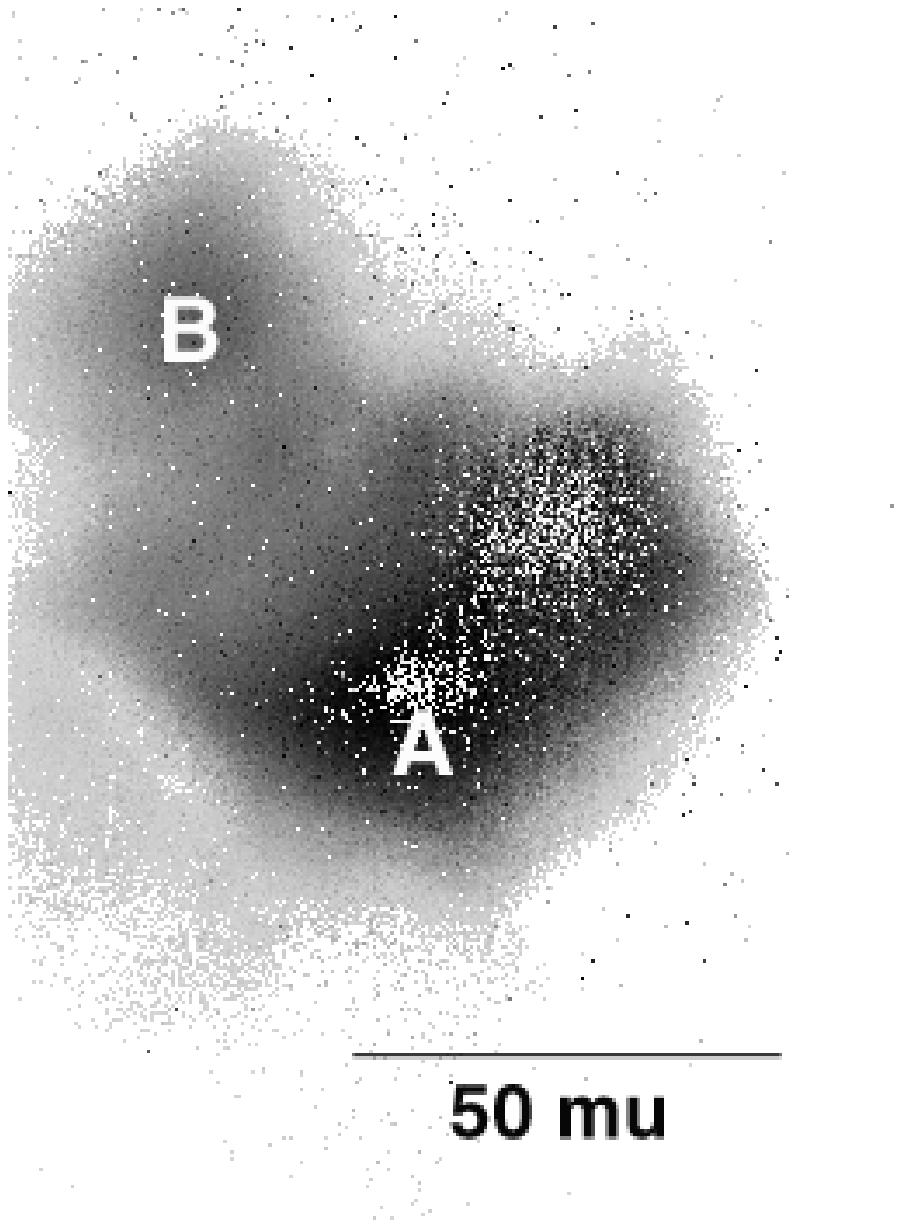,bbllx=5.cm,bblly=8.cm
,bburx=16.cm,bbury=19.cm,width=0.47\textwidth}} & \subfigure[ ]
{\epsfig{figure=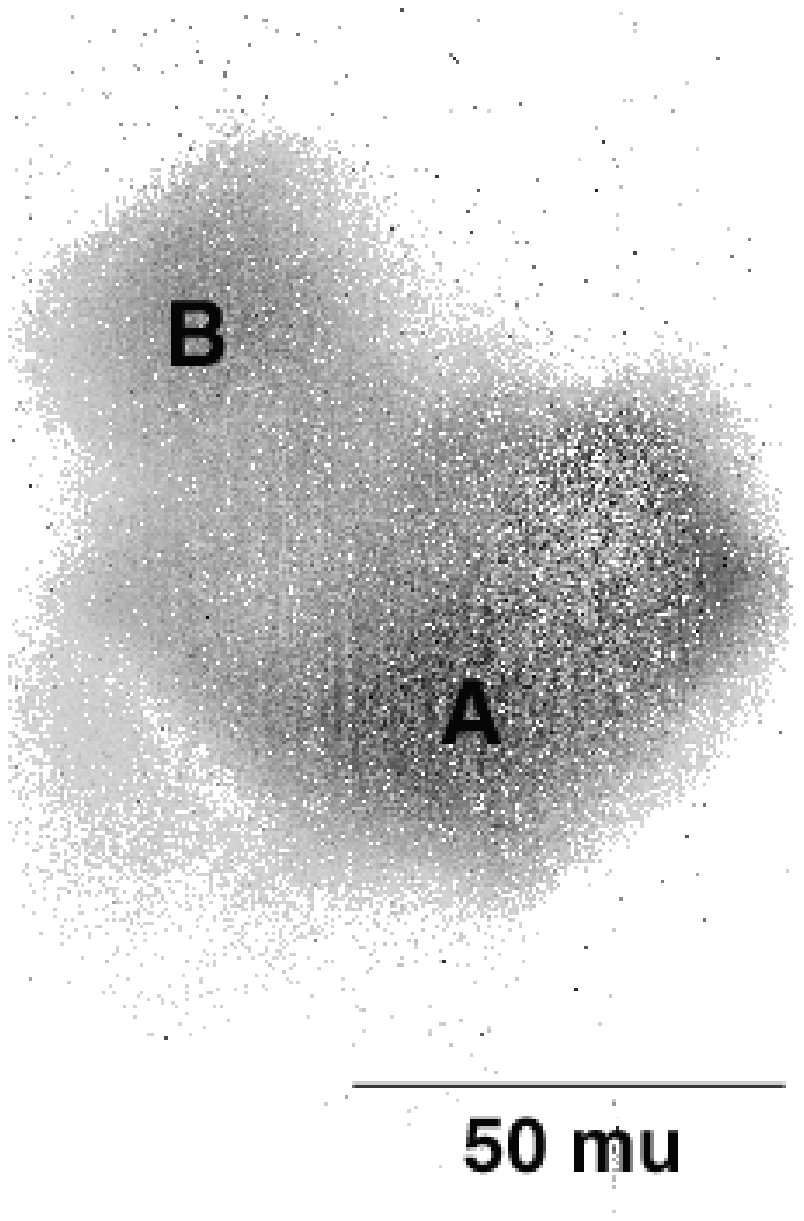,bbllx=5.cm,bblly=8.cm
,bburx=16.cm,bbury=19.cm,width=0.47\textwidth}}
  \end{tabular}
 \end{center}
\caption[]{\label{fig-agathe0}Hot spot {\em X} on the growth side at
field strength of (a) \mbox{$E=5.7{\kV / \cm}$} and (b)
\mbox{$E=4.7{\kV / \cm}$}.}
\end{figure}

\begin{figure}[tmb]
 \begin{center}
  \begin{tabular}{cc}

\subfigure[ ] {\epsfig{figure=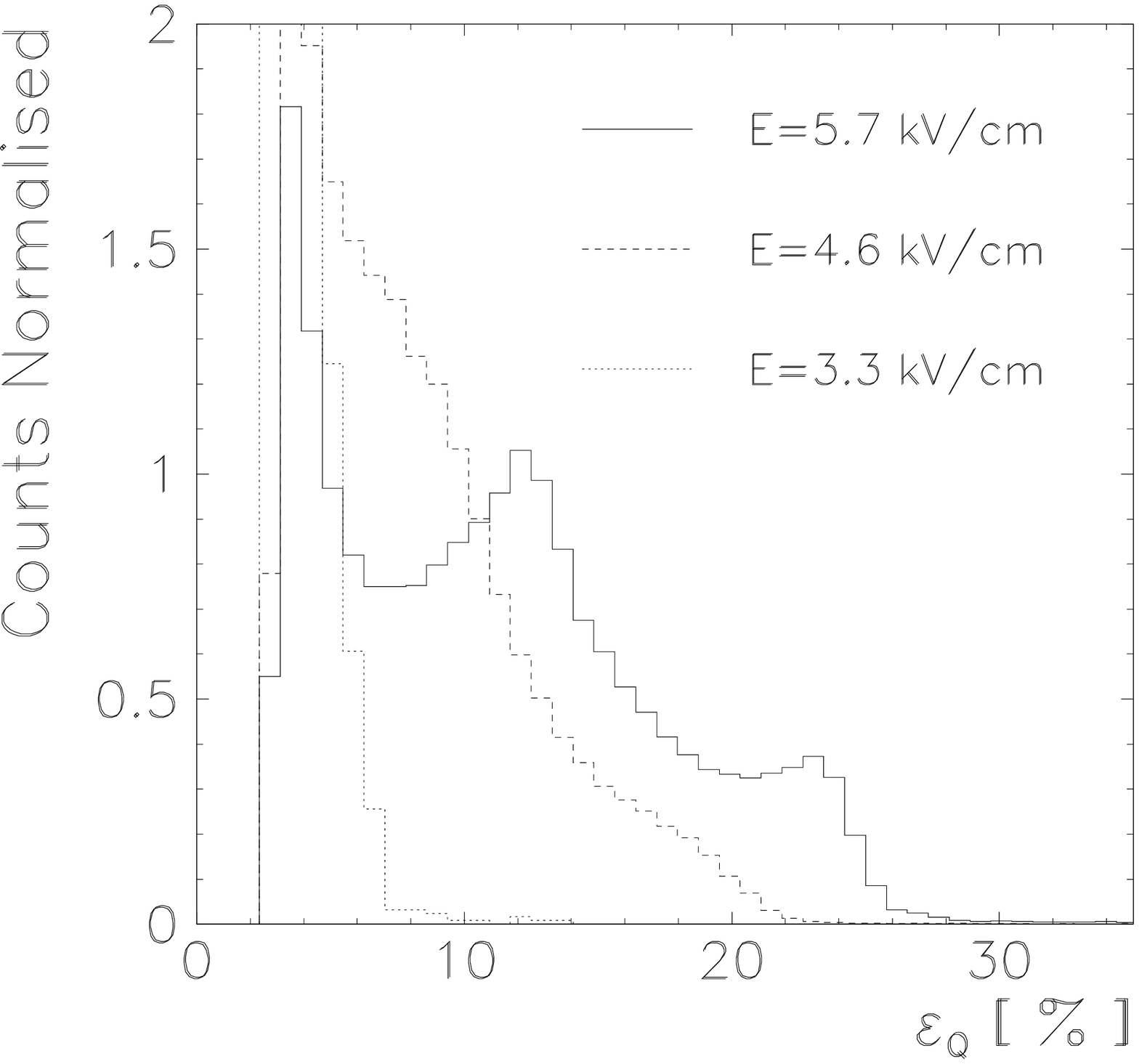,bbllx=0.cm,bblly=0.cm
,bburx=20cm,bbury=20cm,width=0.47\textwidth}} &

\subfigure[ ] {\epsfig{figure=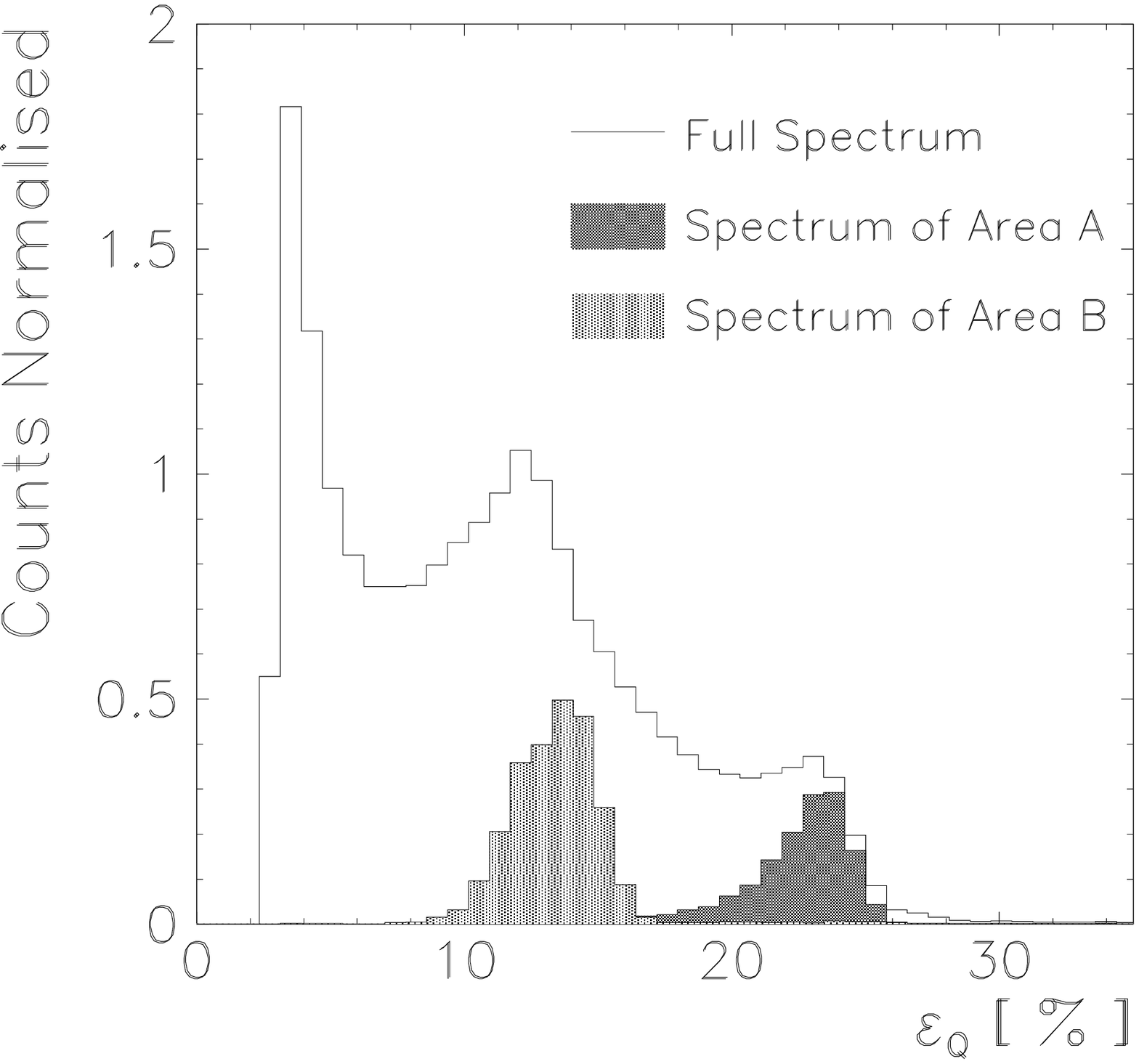,bbllx=0.cm,bblly=0.cm
,bburx=20.cm,bbury=20.cm,width=0.47\textwidth}}

  \end{tabular}
 \end{center}
\caption[]{\label{fig-agathe-AB-spectra}(a) Spectra 
of hot spot {\em X} described
in the text at different electrical field strengths and (b) spectra of
area A and B and full spectrum at $E=5.7 {\kV /\cm}$.}
\end{figure}

The same area of the diamond was scanned with a SEM and the SEM
picture was correlated with the proton micro-beam measurement, as
shown in fig.~\ref{fig-pix-sem}. Due to alignment and scaling
uncertainties we estimate a matching resolution of $100\mum$.
Hot spots were found preferentially on cloven structures between
grains.  A remarkable large hot spot (see the arrow in
fig.~\ref{fig-pix-sem}), which was stable under continuous
irradiation, was located in the middle of the metallisation strip.
The structure of this single hot spot $X$ is displayed in
fig.~\ref{fig-agathe0} for different field strengths.  The counting
rate was reduced when the applied field was lowered and
the corresponding spectra (fig.~\ref{fig-agathe-AB-spectra}a) show a
decrease in pulse height for lower fields. Two distinct peaks in the
spectrum are observed, which are ascribed to two different areas in
the hot spot.  Fig.~\ref{fig-agathe-AB-spectra}b shows the full
spectrum of the hot spot and in shaded the spectra of area A and B at
$E=5.7~{\kV/\cm}$.  Both spectra show clearly separated peaks. Area B
is less efficient than area A. The peak position of A corresponds to a
collection efficiency of about 25\%.

\begin{figure}[tmb]
\begin{center}
\mbox{\psfig{file=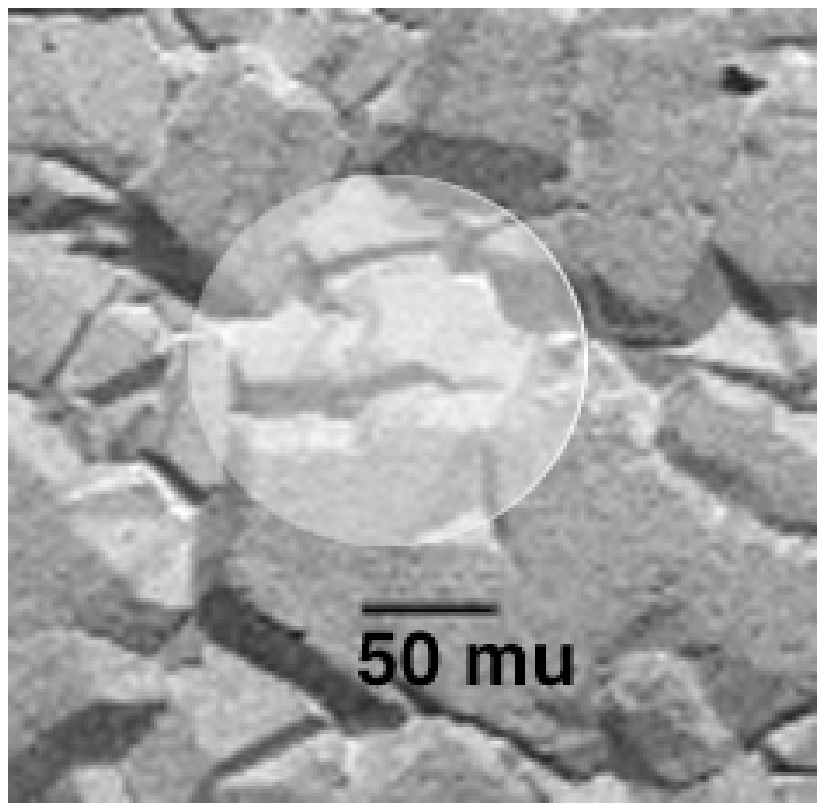,bbllx=190pt,bblly=281pt
,bburx=422pt,bbury=510pt,width=0.47\textwidth}}
\caption[]{\label{fig-agathe-sem}Region of the crystal 
where the hot spot is located. }
\end{center}
\end{figure}

Within the accuracy of the matching of the SEM-picture and the
$\epsilon_{\mathrm{Q}}$ map we assigned the highlighted area shown in
fig.~\ref{fig-agathe-sem} to this hot spot. The size of the
highlighted area includes the estimated error on the matching.  It
shows that the hot spot sits most likely on the cloven structure as
indicated in the picture.

We also scanned the substrate side of the diamond.  The results of the
substrate side gave a similar picture. Again the edges of the
metallisation showed significant higher counting rates and hot spots
appeared also on the substrate side. However, the population with hot
spots was denser on the substrate side (spot density of about
$2.5\mm^{-2}$) and they were smaller in dimension (typically $5\mum$
to $20\mum$) than on the growth side. This can be explained by the
fact that the average grain size is smaller on the substrate side than
on the growth side~\cite{ref-yarbrough}.  Furthermore, the existence
of hot spots on the substrate side indicates that the surface
roughness is not responsible for this phenomenon, since the substrate
side was polished.

We tried to correlate $\epsilon_{\mathrm{Q}}$ maps of the growth side
and the substrate side.  If hot spots would appear at the same
locations on the growth and on the substrate side it could indicate
that continuous crystal columns of high collection efficiency are
present at this locations. However, we could not observe a spatial
correlation between the hot spot from the substrate and the growth
side. This indicates that the lattice orientation or the crystal
quality of the columnar structured crystallites are not a sufficient
condition for a hot spot.  The instability of some hot spots indicate
a close link to polarisation phenomena as proposed by
\cite{ref-manfredotti96}.  The result indicates that the crystallite
orientation, quality and size are not sufficient conditions to produce
a hot spot.  The results favour the interpretation of a complex
polarisation phenomenon at grain boundaries~\cite{ref-manfredotti96}.
This view is supported by the spatially resolved measurement of
$\epsilon_{\mathrm{Q}}$ with 10 keV photons, which ionise uniformly in
the bulk and suppress polarisation.  No enhancement of
$\epsilon_{\mathrm{Q}}$ was seen with photons at the hot spot
locations found with protons.

\section{Summary and interpretation of the result}
\label{s-conclusion}
We have presented four experiments using different types of ionising
radiation to investigate the charge collection properties of CVD
diamond. In particular, we have used beta-particles of up to
$2.28\MeV$, a narrow beam of $10\keV$ photons from a synchrotron
source, $5.8\MeV$ alpha-particles and a $2\MeV$ proton micro-beam.

The samples show an increase of the charge collection efficiency,
$\epsilon_{\mathrm{Q}}$, when irradiated with beta-particles.  The
radiation dose needed to achieve saturation of this so called priming
effect is different by a factor 15 between the two investigated
samples. The relative improvements of $\epsilon_{\mathrm{Q}}$ ($1.63
\pm 0.04 $ for sample A and $1.64 \pm 0.06 $ for sample B) are
comparable.

To obtain information about the lateral distribution of
$\epsilon_{\mathrm{Q}}$, the signal spectrum recorded with beta
particles was fitted with a smeared Vavilov-distribution which results
in an intrinsic broadening of $\epsilon_{\mathrm{Q}}$ of
$\sigma_{\mathrm{intr}}=(4.0^{+0.3}_{-0.7})\%$ for sample A (see
section~\ref{s-mips}).  This value is in good agreement with a direct
measurement of the width and the spatial distribution of
$\epsilon_{\mathrm{Q}}$ with a beam of $10\keV$ photons, which
measures for the same sample $\sigma_{\mathrm{intr}}=(4.32 \pm
0.22)\%.$

In contrast, the response behaviour of the CVD diamond samples to
mono-energetic alpha-particle and protons with stopping ranges of a
few microns indicates a much higher value for
$\sigma_{\mathrm{intr}}$.  The particles produce a very broad pulse
distribution with an exponential fall-off towards large pulse-heights.
A strong polarisation build-up decreases the average value of
$\epsilon_{\mathrm{Q}}$ and the counting rate during irradiation.
This confirms observations reported in~\cite{ref-manfredotti94}.  The
spatially resolved measurement of $\epsilon_{\mathrm{Q}}$ with a
proton micro-beam reveals single spots (hot spots) in the material
with a high response embedded in passive areas, which is in agreement
with observation made in~\cite{ref-oh_dipl95,ref-manfredotti96}.  The
shape of the hot spots and the correlation with SEM-pictures indicates
that hot spots are bound to the crystallite structure.  Since hot
spots appear on the growth and on the polished substrate side we
conclude that hot spots are neither caused by distinct surface
topologies, nor require a certain crystallite size.

The most likely explanation seems to be that particular configurations
of grain boundaries in the bulk, which are most likely to possess a
high trap density~\cite{ref-manfredotti96}, influence the polarisation
field in the neighbourhood of single crystallites.  The unstable
behaviour of some hot spots and the good response on metallisation
edges corroborates this assumption. As a result of this proposed
complex and non-uniform polarisation, $\epsilon_{\mathrm{Q}}$ as seen
by low energy particles is inhomogeneous. This is supported by the
measurements with alpha-particles.  The pulse distributions obtained
with alpha-particles are compatible with the proton micro-beam
measurements.  Furthermore, the hot spots seen with the proton
micro-beam do not show significantly higher values of
$\epsilon_{\mathrm{Q}}$ compared to a $10\keV$ photon beam.  This
suggests that the existence of hot spots is bound not only to a good
crystal quality in this particular spot, but also to distinct features
of the polarisation field in this area.  The maps of
$\epsilon_{\mathrm{Q}}$ obtained with a proton micro-beam are highly
distorted by the local polarisation fields. Thus this method is not
suitable to determine maps of $\epsilon_{\mathrm{Q}}$ for MIPs.

The observed increase of $\epsilon_{\mathrm{Q}}$ under irradiation
with MIPs and the decrease of $\epsilon_{\mathrm{Q}}$ observed under
irradiation with protons and alpha particles is not contradictory.  In
the former case traps are filled throughout the bulk thus increasing
$\epsilon_{\mathrm{Q}}$ due to the decrease of active trap
levels. However, because of the rather homogeneous distribution of the
trapped charge no effective polarisation field is produced. In the
latter case the majority of the charge is deposited in a shallow
layer. In the electric field the charge carriers are separated and
trapped and thus create a strong polarisation field. The polarisation
is dominating over the increase of $\epsilon_{\mathrm{Q}}$ from trap
passivation.

The applicability of CVD diamond as a particle detector depends on the
radiation to be detected. For radiation which deposits charge
inhomogeneously as in the case of alpha-particles, whose stopping
range is much less than the film thickness, $\epsilon_{\mathrm{Q}}$
varies strongly and the stability is very poor. For homogeneous
ionisation typically produced by MIPs or synchrotron-radiation at
photon energies with $\lambda_0 \gg d$, $\epsilon_{\mathrm{Q}}$ is
stable if the detector is in its primed state, and a Gaussian-like
distribution is observed.  Thus, we conclude that CVD diamond has the
potential of being used as a stable detector for homogeneous
ionisation density distributions. However the relatively high value of
$\sigma_{\mathrm{intr}}$ might limit the achievable resolution of a
position sensitive device.

\begin{ack}
We would like to thank Prof.~Dr.~Dr.~h.~c.~G.~Lindstr\"om and
Dr.~E.~Fretwurst from the II.~Institut f\"ur Experimentalphysik of the
Univerit\"at Hamburg for their support of the measurements with alpha
particles and of the I-V-curves, Dr.~E.~Fretwurst and Dr.~M.~Niecke
for their support of the measurements with a proton micro-beam, and
Dr.~T.~Wroblewski from HASYLAB for his support of the measurements at
the synchrotron facility. We would also like to thank A.~Bluhm and
Dr.~L.~Sch\"afer from the Fraunhofer Institut f\"ur Schicht- und
Oberfl\"achentechnik, Braunschweig, for the preparation of sample B
and M.~Zeitler of the Univesit\"at Augsburg for the pole figure
measurements.
\end{ack}


\begin{thebibliography}{99}
\bibitem{ref-fhg_metal} {Metallisations were carried out
at~\cite{ref-fhg_addr}.}
\bibitem{ref-blabla} {The measurements with a proton micro-beam were
carried out at the I. Institut f\"ur Experimentalphysik, University of
Hamburg.}
\end{thebibliography}
\end{document}